\documentclass[a4paper,11pt]{article}
\pdfoutput=1

\usepackage{jcappub} 
\usepackage{graphicx}
\usepackage{dcolumn}
\usepackage{bm}

\usepackage[english]{babel}
\usepackage[latin1]{inputenc} 

\usepackage{epstopdf}
\usepackage{amsfonts}
\usepackage{color}
\usepackage{epsfig}
\usepackage{mathrsfs}
\usepackage{amsmath,amssymb}
\usepackage{subfigure}

\usepackage{framed}
\usepackage{lipsum}
\usepackage[svgnames]{xcolor}
\definecolor{shadecolor}{named}{LightGrey}

\usepackage{float}
\floatstyle{plaintop}
\restylefloat{table}
\usepackage[tableposition=top]{caption}

\def\bx{\mathbf{x}}
\def\bz{\mathbf{z}}
\def\bU{\mathbf{U}}
\def\bX{\mathbf{X}}
\def\bW{\mathbf{W}}
\def\bZ{\mathbf{Z}}
\def\bdelta{\mbox{\boldmath $\delta$}}

\def\bmu{\mbox{\boldmath $\mu$}}
\def\bphi{\mbox{\boldmath $\phi$}}
\def\btheta{\mbox{\boldmath $\theta$}}
\def\bSigma{\mbox{\boldmath $\Sigma$}}

\def\cL{\mathcal{L}}

\def\sbX{{\underset{\sim}{\bX}}}
\def\sbZ{{\underset{\sim}{\bZ}}}

\DeclareMathOperator\erf{erf}
\def\Xmax{\ifmmode {X_\mathrm{max}}\else
                   {$X_\mathrm{max}$}\fi\xspace}%
                   
\title{\boldmath Modeling high energy cosmic rays mass composition data via mixtures of multivariate skew-t distributions}

\author[a]{S. Riggi,\footnote{Corresponding author.}}
\author[b]{S. Ingrassia}

\affiliation[a]{INAF - Osservatorio Astrofisico di Catania, Italy}
\affiliation[b]{Department of Economics and Business, University of Catania, Italy}

\emailAdd{simone.riggi@ct.infn.it}
\emailAdd{s.ingrassia@unict.it}

\abstract{
We consider multivariate skew-t distributions for modeling composition data of high energy cosmic
rays. The model has been validated with simulated data for different primary nuclei and hadronic models focusing on the depth of maximum 
$X_\mathrm{max}$ and number of muons $N_{\mu}$ observables. Further, we consider mixtures of multivariate skew-t distributions for cosmic 
ray mass composition determination and event-by-event classification. With respect to other approaches in the field,
it is based on analytical calculations and allows to incorporate different sets of constraints provided by the present hadronic models. We present some
applications to simulated data sets generated with different nuclear abundances assumptions.
As it does not fully rely on the hadronic model predictions, the method is particularly suited to the current experimental scenario in which 
evidences of discrepancies of the measured data with respect to the models have been reported for some shower observables, such as the number of muons at ground level.
}

\begin{document} 
\maketitle
\flushbottom

\section{Introduction}\label{IntroductionSection}
The nuclear composition of cosmic ray particles is a crucial observable to explore the origin of such radiation at energies above $10^{18}$ eV. 
Different theoretical models have been advanced in this direction predicting a given flux from each nuclear component. The measurement of the relative abundances 
as a function of the primary energy at least for groups of nuclear components therefore provide a deep insight in the field and contribute to significantly constrain the 
present theories. 

At energies above the knee the measurement of the mass must necessarily be based on indirect techniques, making use of shower parameters sensitive to the primary mass. 
Among these, the depth $X_\mathrm{max}$ at which the longitudinal development has its maximum and the number of muons $N_{\mu}$ reaching ground level 
are the most powerful observables. A recent review of the mass composition methods and experimental results over the entire energy spectrum range 
is presented in \cite{UngerKampertReview}.\\At present, different statistical methods are employed depending on whether the composition information has to be
reconstructed on an event-by-event basis. For instance this results to be extremely helpful when studying possible correlations of the shower arrival direction 
with given astrophysical objects, in proton-Air cross section analysis or in the search of gamma ray or neutrino events in the hadronic background. In all 
these situations pattern recognition methods, e.g. neural networks as in \cite{Tiba,Ambrosio,RiggiNIMPreparation}, linear 
discriminant analysis \cite{NeutrinoAugerPaper}, supervised clustering algorithms as in \cite{RiggiNIMPreparation}, are typically adopted. 
To determine the nuclear composition on average, binned likelihood methods, as in \cite{DUrsoICRC09}, or unfolding analyses, as in \cite{KASCADEAnalysis}, are preferred. 

The presence of stochastic shower-to-shower fluctuations and the experimental resolution currently achieved in the measurement of the shower parameters impose a limit 
on the number of mass groups to be determined in both kind of analyses, typically less than five.

All these methods directly compare the observed data with mixtures of distributions generated with detailed Monte Carlo simulations of the shower development 
in atmosphere for each nucleus. The only parameters to be optimized are the component weights.

Shower simulations strongly rely on theoretical models 
of the hadronic interactions extrapolated at energies well beyond those reached in modern accelerators. Among them we cite \textsc{Qgsjet01} \cite{QGSJET01Paper}, 
\textsc{QgsjetII} \cite{QGSJETIIPaper}, \textsc{Sibyll} \cite{SibyllPaper} and \textsc{Epos} \cite{EPOSPaper}.
Surprisingly, most of the shower observables measured so far at such extreme energies are well described or at least bracketed by the hadronic model predictions, 
i.e. see \cite{UngerKampertReview}.
Furthermore, the discrepancies among different models will be significantly reduced with incoming model releases accounting for recent LHC measurements, 
for instance see \cite{PierogECRC2012}. Recently, however, a significant discrepancy between measured data and Monte Carlo simulations has been reported in 
the muon number estimator both for vertical \cite{AllenICRC2011} and for very inclined showers \cite{GonzaloICRC2011}. The origin of such discrepancy is currently
under investigation.

This scenario motivates the development of a statistical model of the shower observables and 
an unsupervised approach, incorporating the valid constraints provided by the 
hadronic models, for mass composition fitting and classification purposes. These represent the main goals of the present study, which is organized as follows.

In Section \ref{ModelSection} we explored the possibility of modeling all shower observables, focusing our attention on the depth of maximum 
$X_\mathrm{max}$ and the number of muons $N_{\mu}$, 
by using a multivariate skew-t distribution. This and similar classes of skew distributions are receiving a growing attention in model-based clustering 
over the last few years, i.e. see the works by Lee and Mc Lachlan \cite{LeeMcLachlan2012}, Vrbik and McNicholas \cite{McNicholas2012}, 
Franczak et al \cite{Franczak2012} and Azzalini and Genton \cite{Azzalini2008}, due to their flexibility to model non-gaussian asymmetric data and the possibility 
of developing elegant and relatively computationally straightforward mixture solutions in the EM framework. The validation of the model with Monte Carlo shower
simulations is reported in paragraph \ref{DataSection}.\\In Section \ref{MSTMixtureSection} we describe the clustering algorithm adopted for composition determination.
It is mainly based on the work by McNicholas et al \cite{McNicholas2012} for what concerns the analytical solutions to 
fit mixtures of multivariate skew-t distributions using the EM algorithm. With respect to the latter work we introduced in Section \ref{ConstraintSection} a procedure to 
account for different kind of constraints in the optimization directly derived from shower simulations.
Finally, in Section \ref{AnalysisSection} we applied the method to sample data sets simulated with different nuclear abundance assumptions. The achieved fitting
and classification performances are presented and discussed.

\begin{figure}[!ht]
\centering%
\subtable[p ]{\includegraphics[scale=0.2]{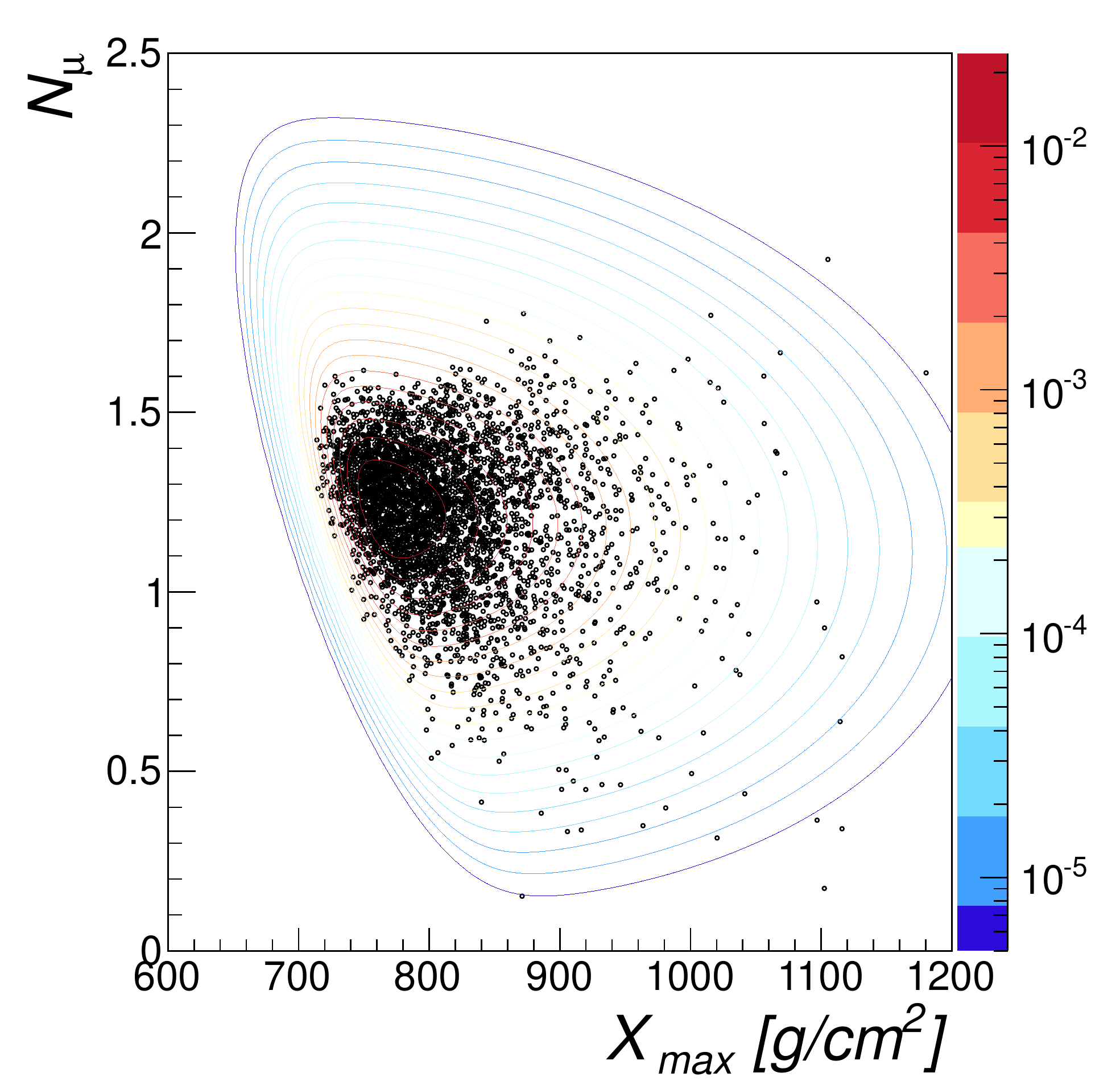}\label{MSFFitFig1}}
\subtable[He ]{\includegraphics[scale=0.2]{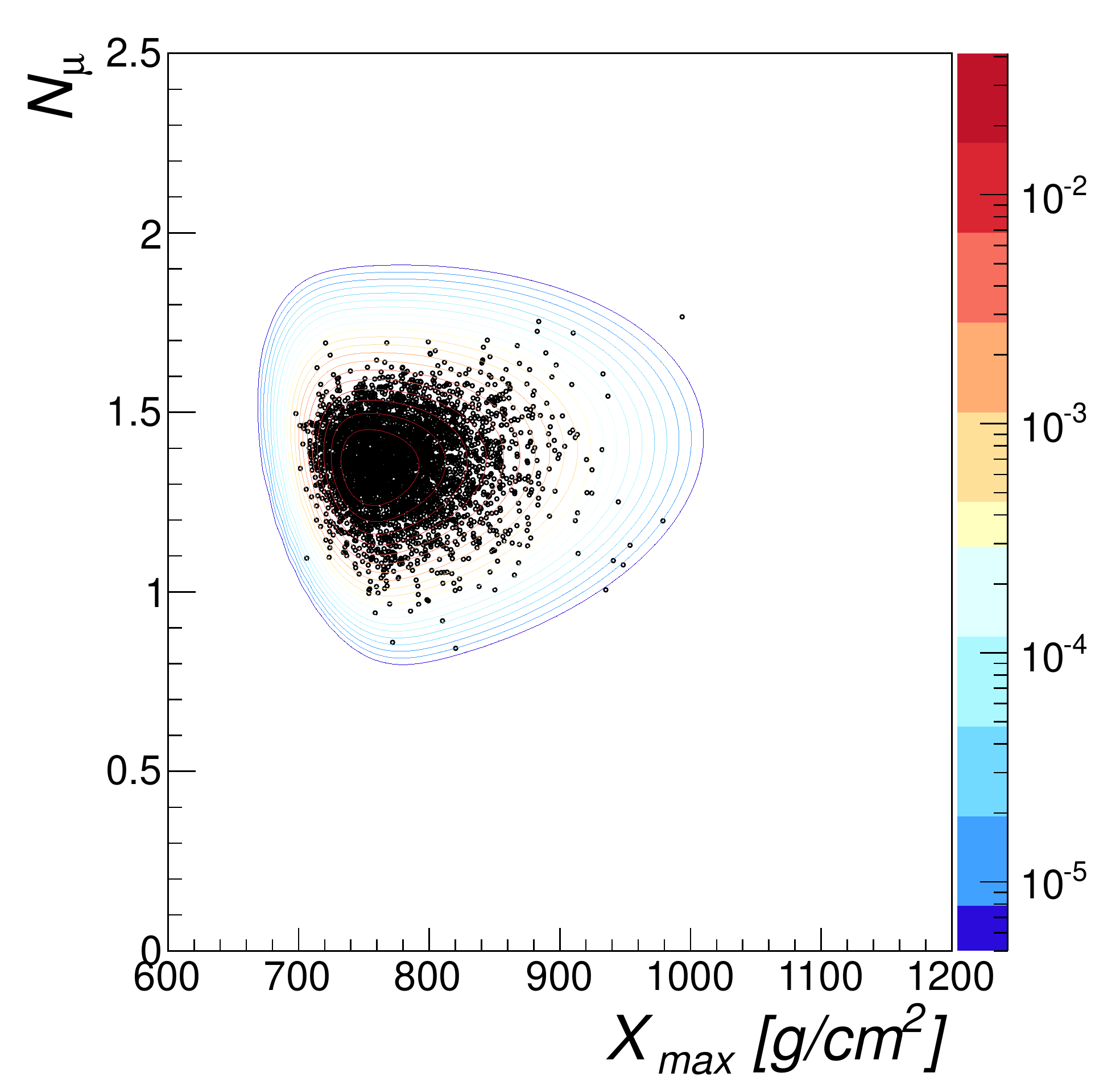}\label{MSFFitFig2}} 
\subtable[Li ]{\includegraphics[scale=0.2]{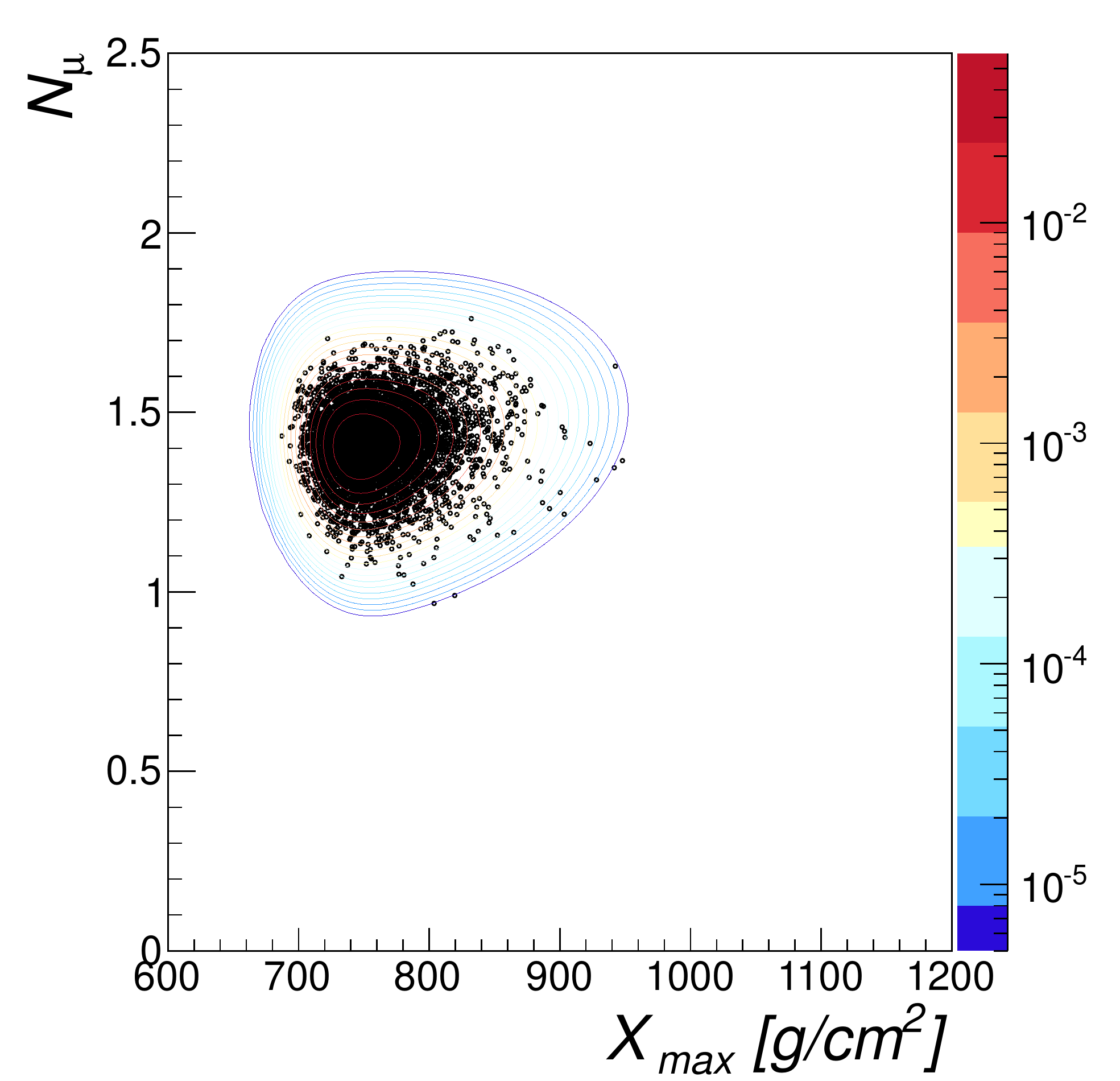}\label{MSFFitFig3}}\\
\vspace{-0.2cm}
\subtable[C ]{\includegraphics[scale=0.2]{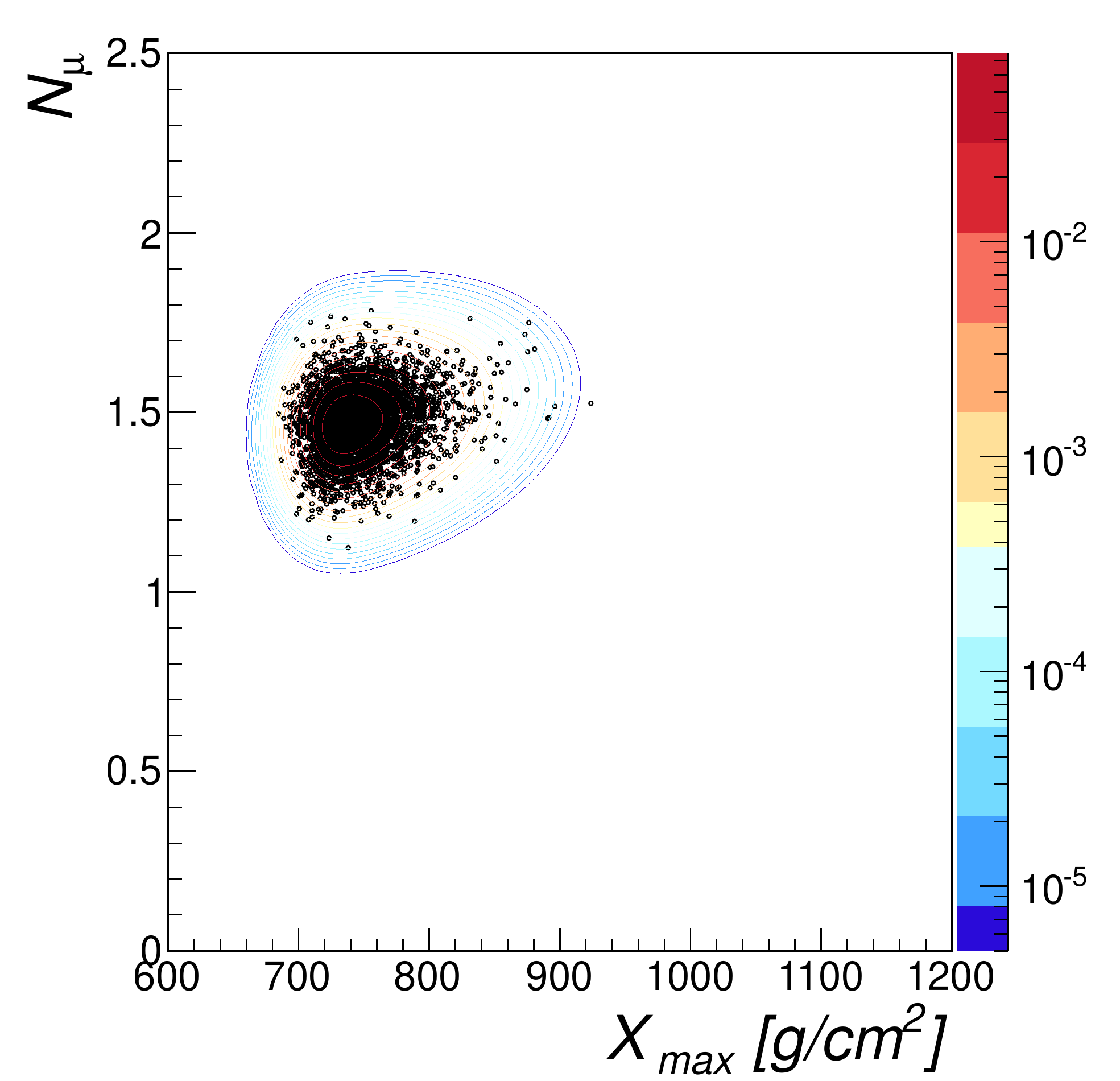}\label{MSFFitFig4}}
\subtable[N ]{\includegraphics[scale=0.2]{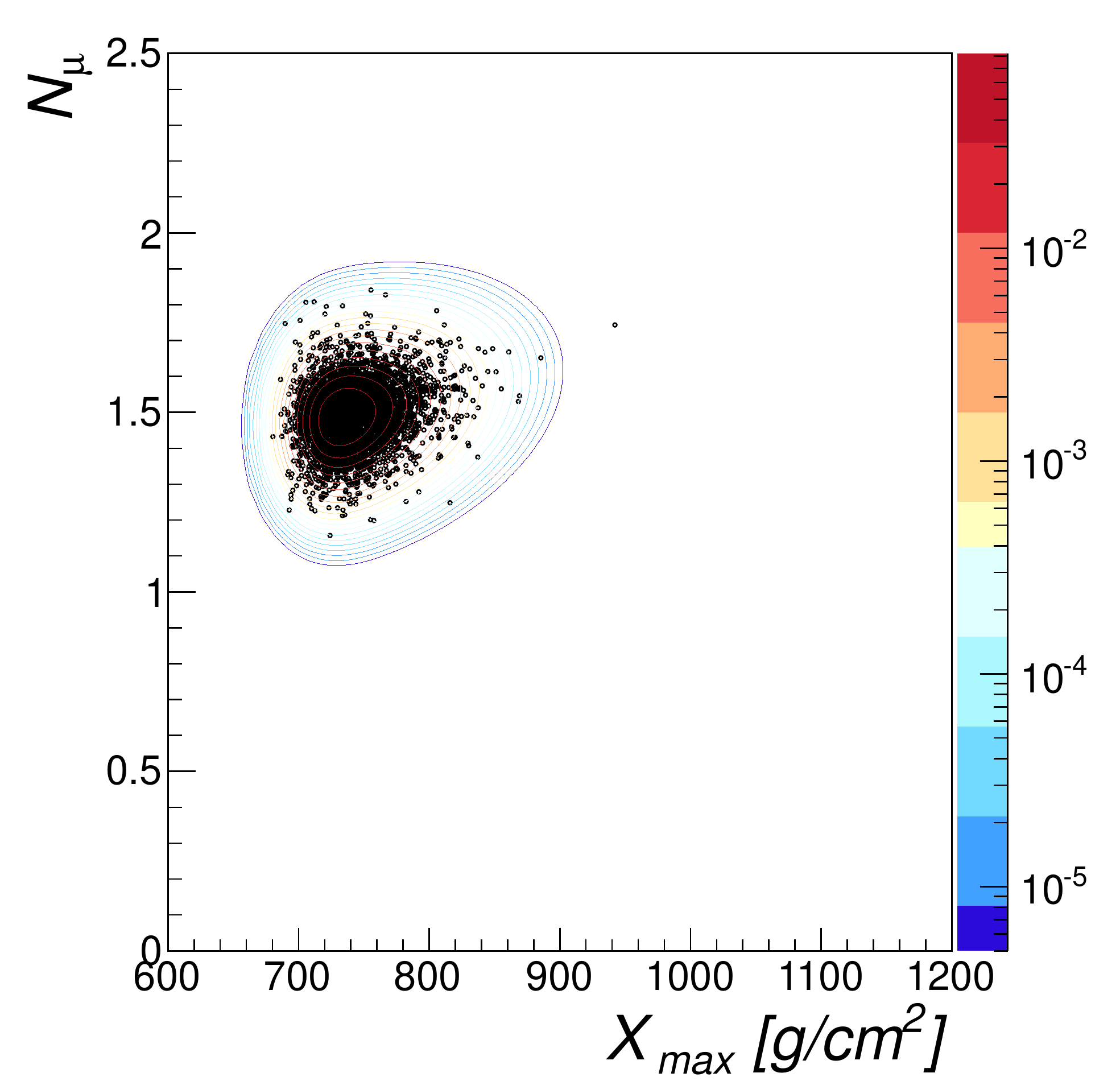}\label{MSFFitFig5}} 
\subtable[O ]{\includegraphics[scale=0.2]{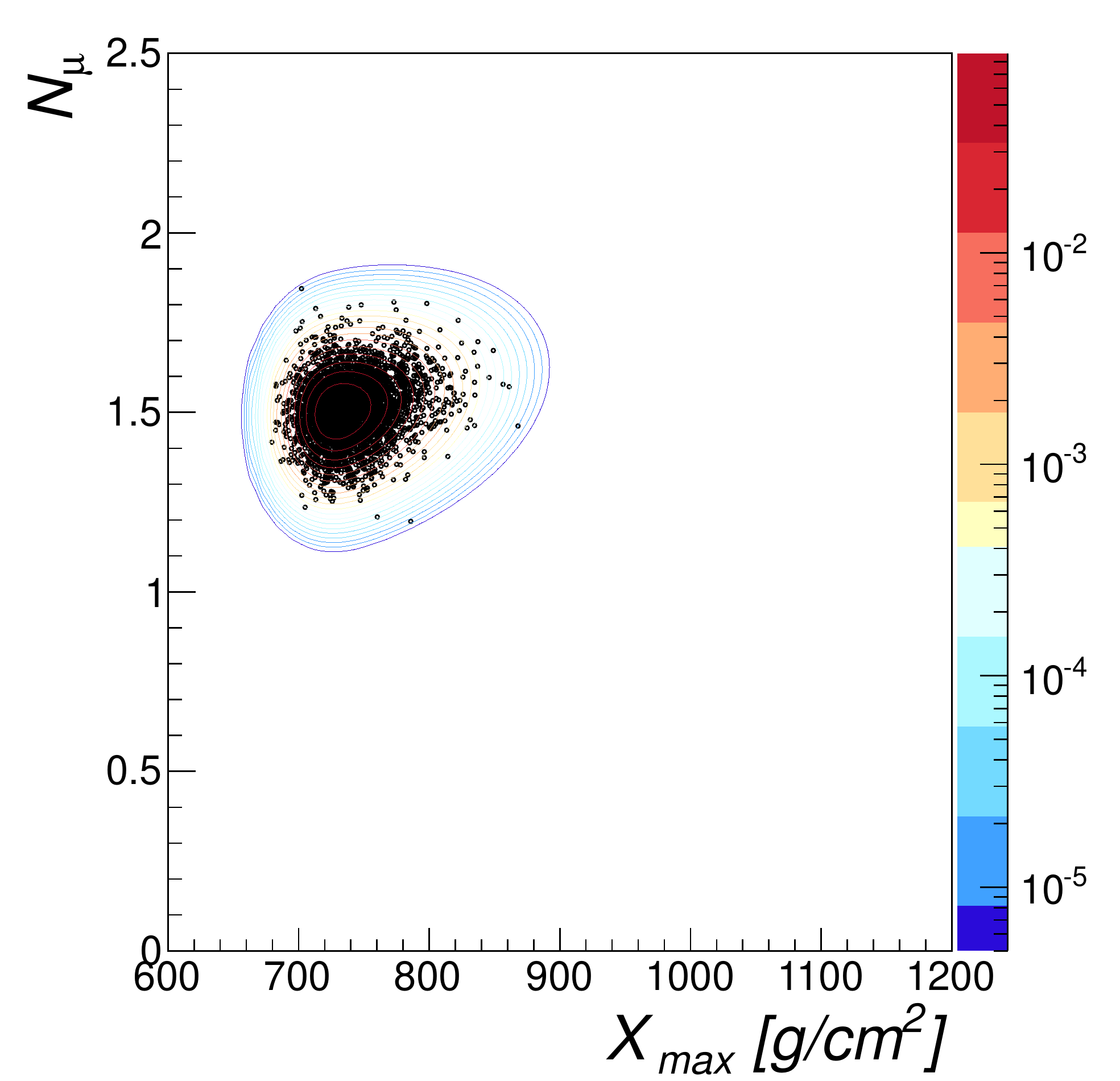}\label{MSFFitFig6}}\\ 
\vspace{-0.2cm}
\subtable[Si ]{\includegraphics[scale=0.2]{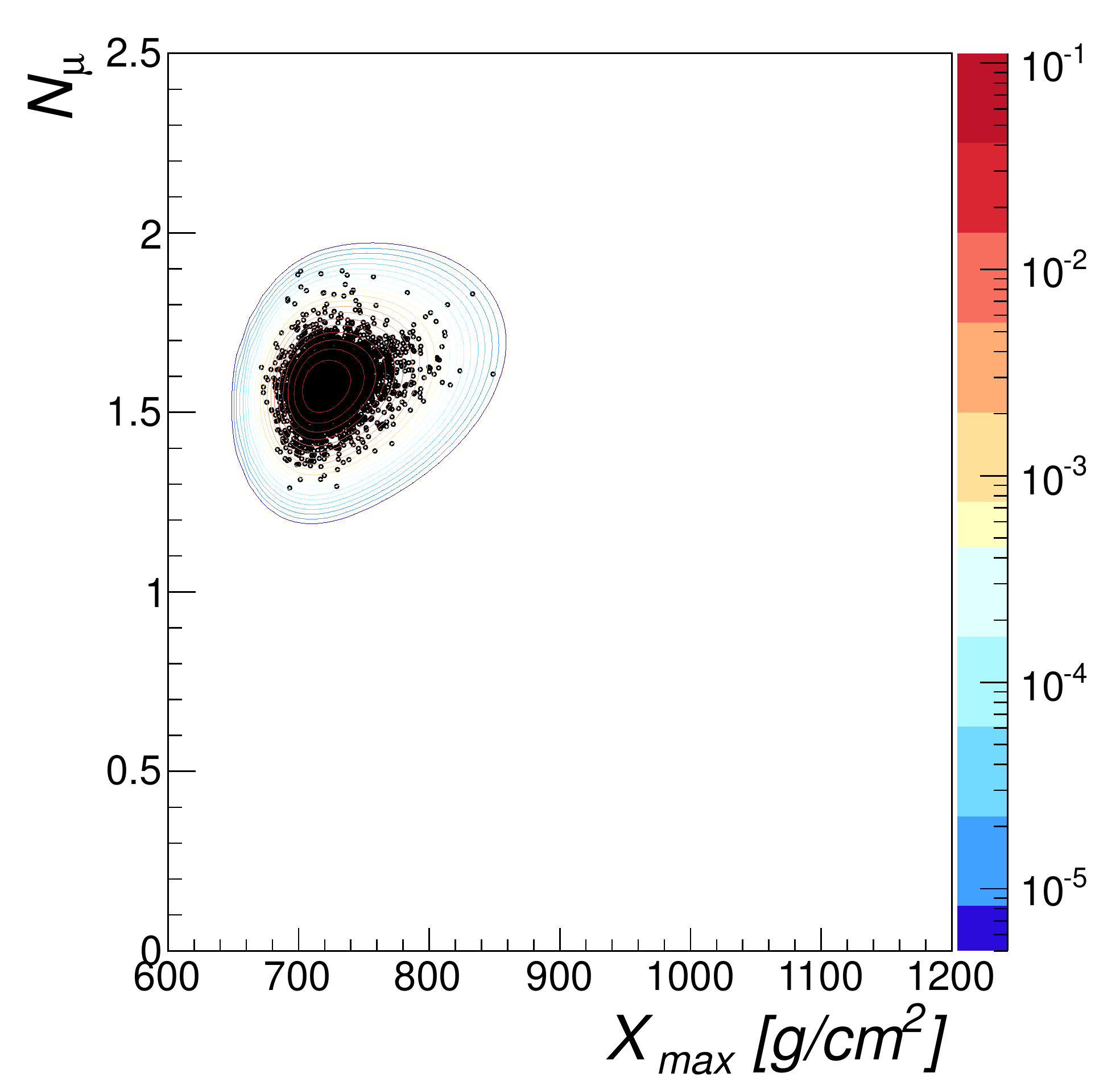}\label{MSFFitFig7}} 
\subtable[Ca ]{\includegraphics[scale=0.2]{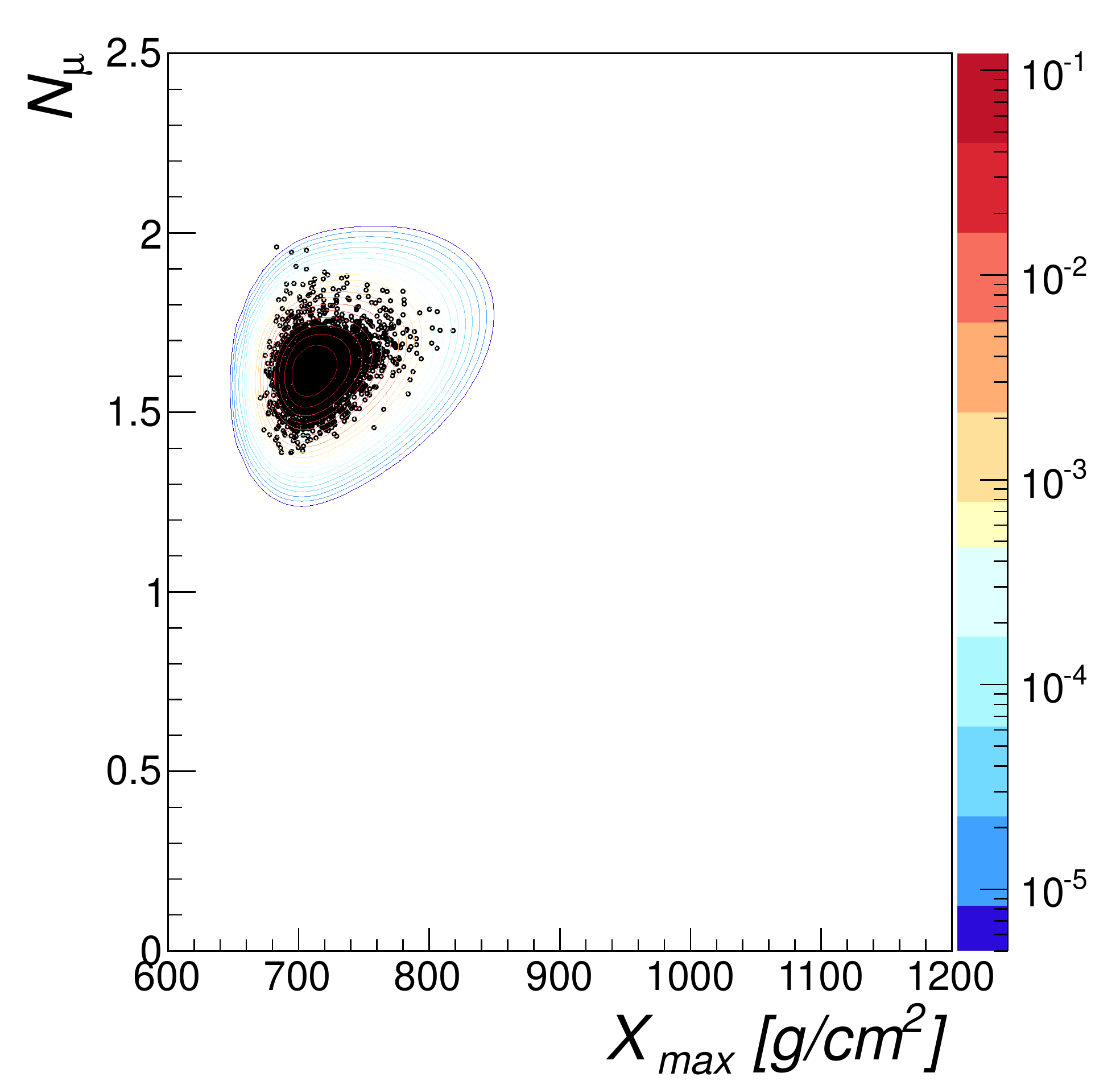}\label{MSFFitFig8}}
\subtable[Fe ]{\includegraphics[scale=0.2]{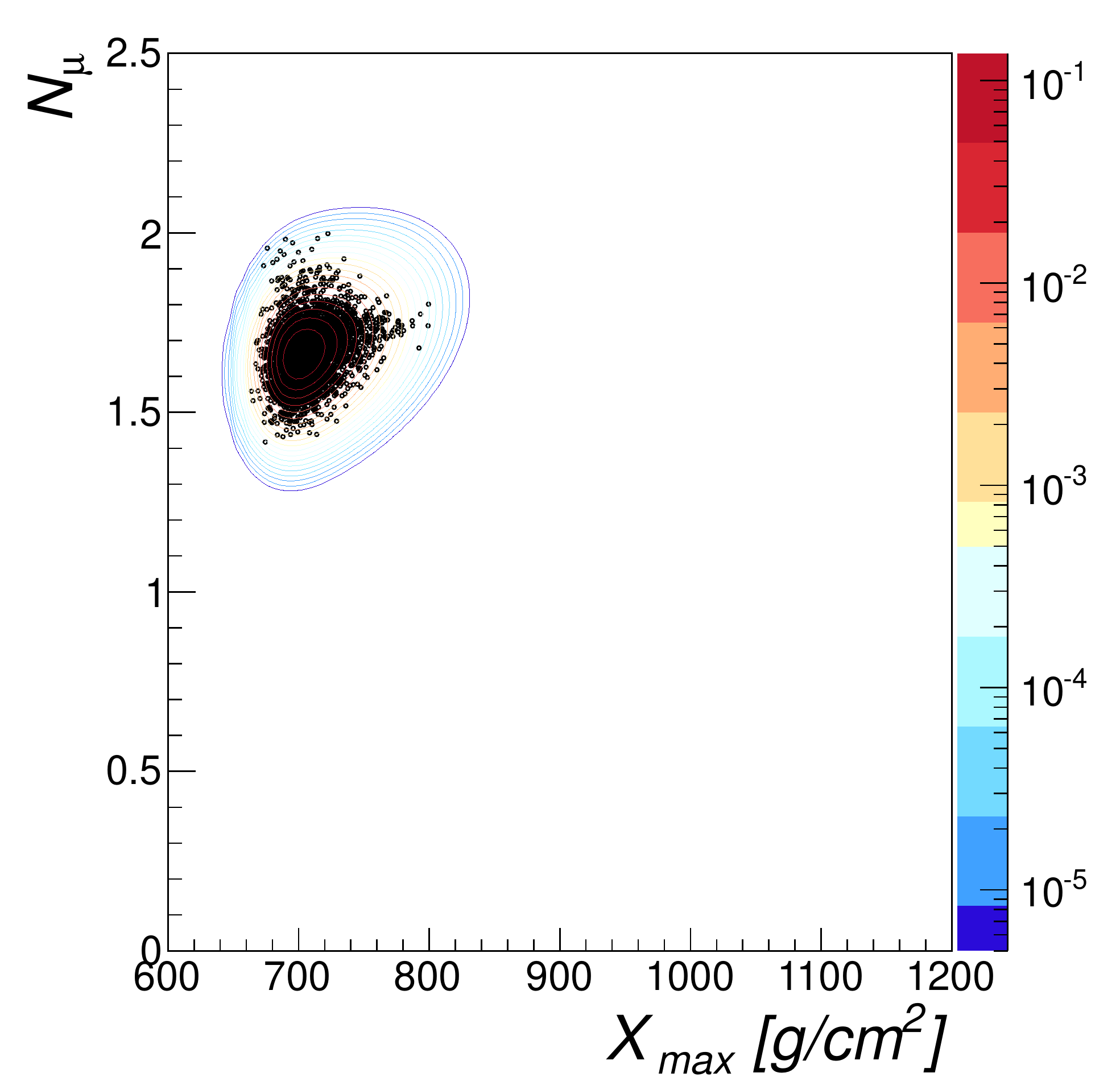}\label{MSFFitFig9}}
\vspace{-0.1cm}
\caption{$X_\mathrm{max}$-$N_{\mu}$ distribution for different primary nuclei at $10^{19}$ eV predicted by the \textsc{Epos}1.99 model.}%
\label{MSTFitFig}
\end{figure}

\section{Modeling the distribution of cosmic ray shower observables}\label{ModelSection}
The fluctuations observed in the shower observables for a given primary nucleus and energy are related to the stocastic fluctuations in the 
position of the first interaction point in the top layers of the atmosphere and in the secondary interactions occuring along the shower development. 
The latter can be considered as Gaussian distributed, given the large number of particles involved, while an exponential distribution is assumed
for the interaction probability. The resulting distribution is asymmetric with a given degree of skewness depending on the considered parameters. 
It turns out from these considerations that a suitable distribution describing the fluctuation of a single shower observable $x$ is a 
convolution of the exponential and gaussian distributions (\textit{exponentially modified gaussian (EMG))} \cite{Ulrich}:

\begin{equation}\label{EMGDefinition}%
f(x)=\frac{1}{2\lambda}\exp{\left(-\frac{x-\mu}{\lambda}+\frac{\sigma^{2}}{2\lambda^{2}}\right)}
\left[1+ \erf\left(\frac{x-\mu}{\sigma\sqrt{2}}-\frac{\sigma}{\lambda\sqrt{2}}\right)\right]
\end{equation}
where $\lambda$ is the attenuation parameter of the exponential function, 
$\mu$ and $\sigma$ the mean and width parameters of the gaussian function, and $erf$ the standard error function.
The above model is based on physical considerations and provides a good representation of both simulated and real data. 
However, to our knowledge, little efforts have been carried out to describe the joint distribution of several shower observables and 
no analytical solutions exists to fit a mixture of EMG distributions. For this reason in this work we propose a different model, based on the 
\textit{multivariate skew-t (MST)} distribution, 
to describe the fluctuations of $m$ shower observables $\bx$. It is as flexible as model \ref{EMGDefinition} in reproducing the skewness 
observed in the shower observables and, as we will show in section \ref{MSTMixtureSection}, it has the great advantage that a closed form 
exists for mixture fitting with the EM algorithm. 

Based on \cite{Pyne2009,Wang2009,McNicholas2012}, we say that a random vector $\bX$ follows a $q$-variate skew-$t$ distribution $f(\bx; \bphi)$ with location vector 
$\bmu$, covariance matrix $\bSigma$, skewness vector $\bdelta$ and $\nu$ degrees of freedom if it can be represented by 
\begin{equation*}
\bX= \bdelta|U|+\bU_{0}
\end{equation*}
where  $\bU_{0}|u, w\sim N_{p}(\bmu+\bdelta|u|,\bSigma/w)$, $U|w\sim N(0,1/w)$ and $W\sim\Gamma(\nu/2,\nu/2)$. The joint distribution of $\bX, U$ and $W$ is given
by:
\begin{equation}
f(\bx,u,w)= Cw^{R-1}e^{-wS} 
\end{equation}
where $\Gamma(\cdot)$ is the Gamma function and:
\begin{align*}
S=&\frac{1}{2}[(\bx-\bmu-\bdelta|u|)^{T}\bSigma^{-1}(\bx-\bmu-\bdelta|u|)+u^{2}+\nu]\\
R=&(\nu+p+1)/2\\
C=& (\nu/2)^{\nu/2}\left[(2\pi)^{\frac{p+1}{2}}|\bSigma|^{1/2}\Gamma(\nu/2)\right]^{-1} \, .
\end{align*}
Integrating out $u$ and $w$ from the joint distribution we get the expression of the multivariate skew-t distribution:
\begin{equation}\label{MSTDefinition}
f(\bx; \bphi)= 2C\Gamma(R)\frac{D^{\frac{1}{2}-R}}{\sqrt{a}}I_{1}\left(R,\frac{b}{\sqrt{Da}}\right) 
\end{equation}
where:
\begin{align*}
a & = \frac{1}{2}(1+\bdelta' \bSigma^{-1}\bdelta)\\
b & = -\frac{1}{2}(\bx-\bmu)' \bSigma^{-1}\bdelta\\
c & =\frac{1}{2}(\bx-\bmu)'\bSigma^{-1}(\bx-\bmu)+\nu/2\\
D& = c-b^{2}/a%
\end{align*}
\begin{align*}
I_{1}(E,\alpha)=\int_{\alpha}^{\infty}(1+x^{2})^{-E}dx=&\frac{\pi\Gamma(2E-1)}{\Gamma(E)^{2}2^{2E-1}}-\alpha \;  _2F_{1}\{1/2,E,3/2,-\alpha^{2}\}
\end{align*}
in which $_{2}F_{1}$ denotes the generalized Gauss hypergeometric function, given by:
\begin{equation*}
_2F_1 (x,y; z; 1) = \frac{\Gamma(z) \Gamma(z-x-y)}{\Gamma(z-x) \Gamma(z-y)} \, .
\end{equation*}

Finally, $\bphi$ denotes the overall parameter of the distribution, that is $\bphi=(\bmu, \bSigma, \bdelta, \nu)$.
\begin{figure}[!t]
\centering
\subtable[$\mu$ constraints]{\includegraphics[scale=0.25]{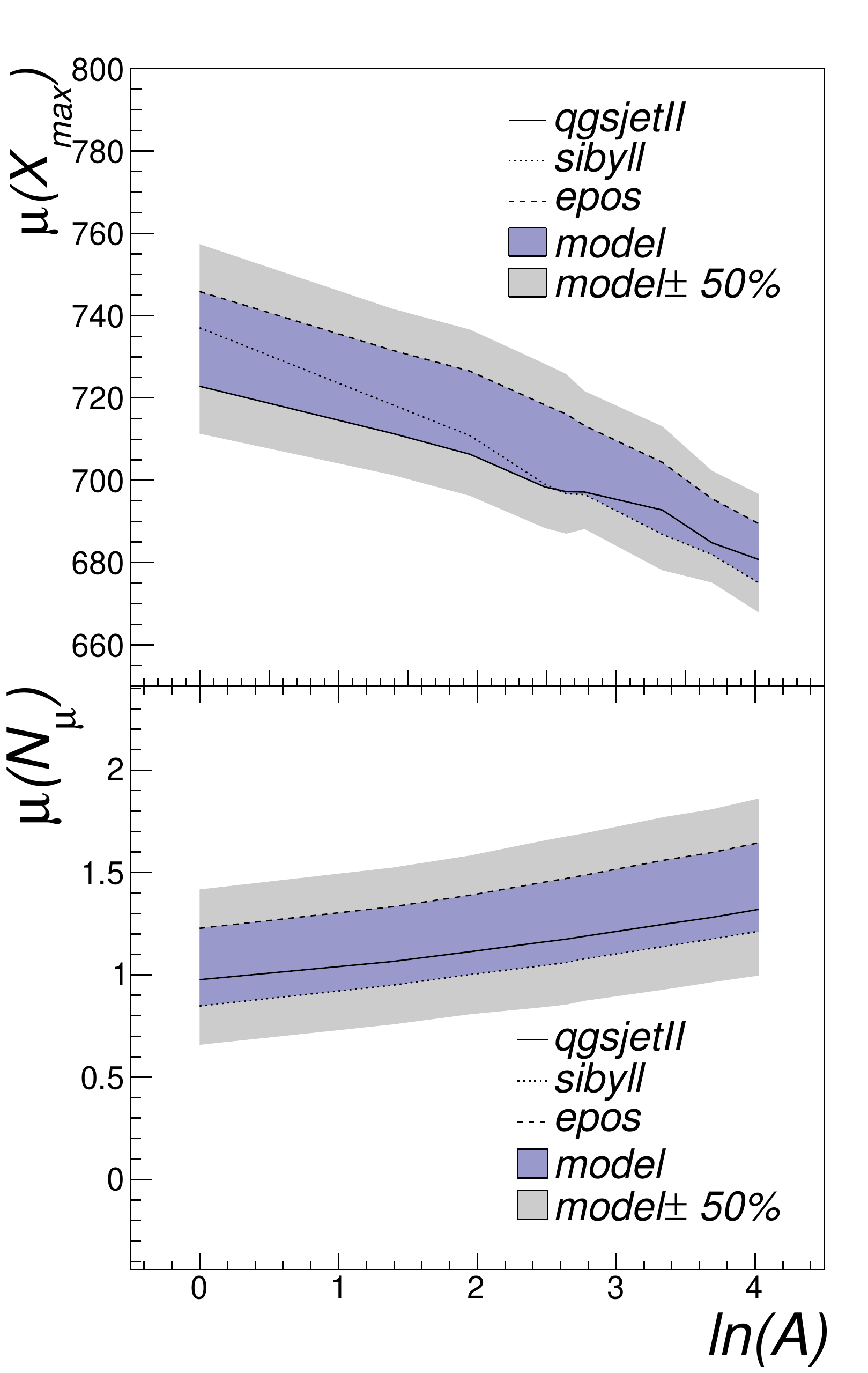}\label{MeanConstraintFig}}
\hspace{-0.4cm}
\subtable[$\Sigma$ constraints]{\includegraphics[scale=0.25]{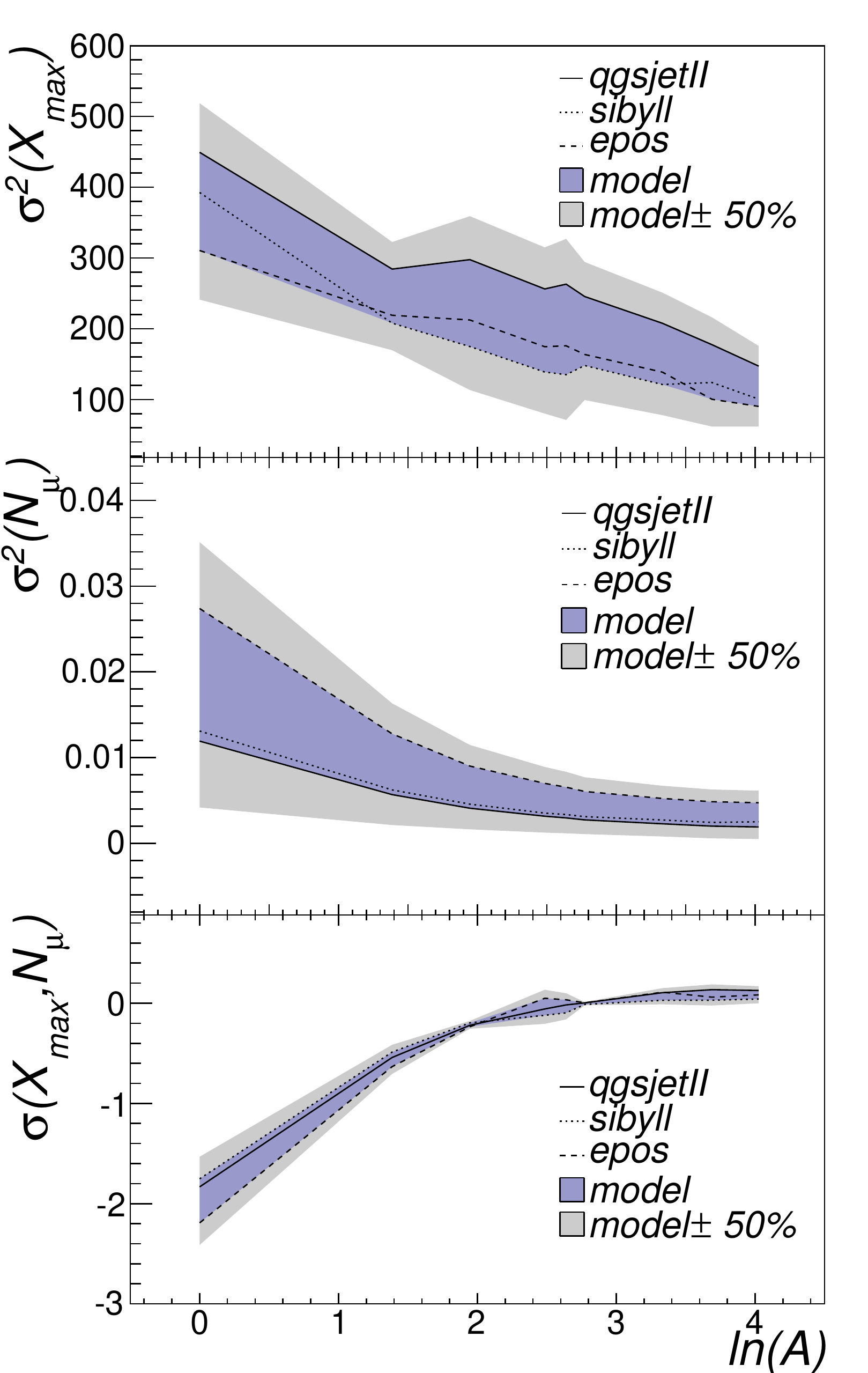}\label{CovarianceConstraintFig}}
\hspace{-0.3cm}%
\subtable[$\delta$ constraints]{\includegraphics[scale=0.25]{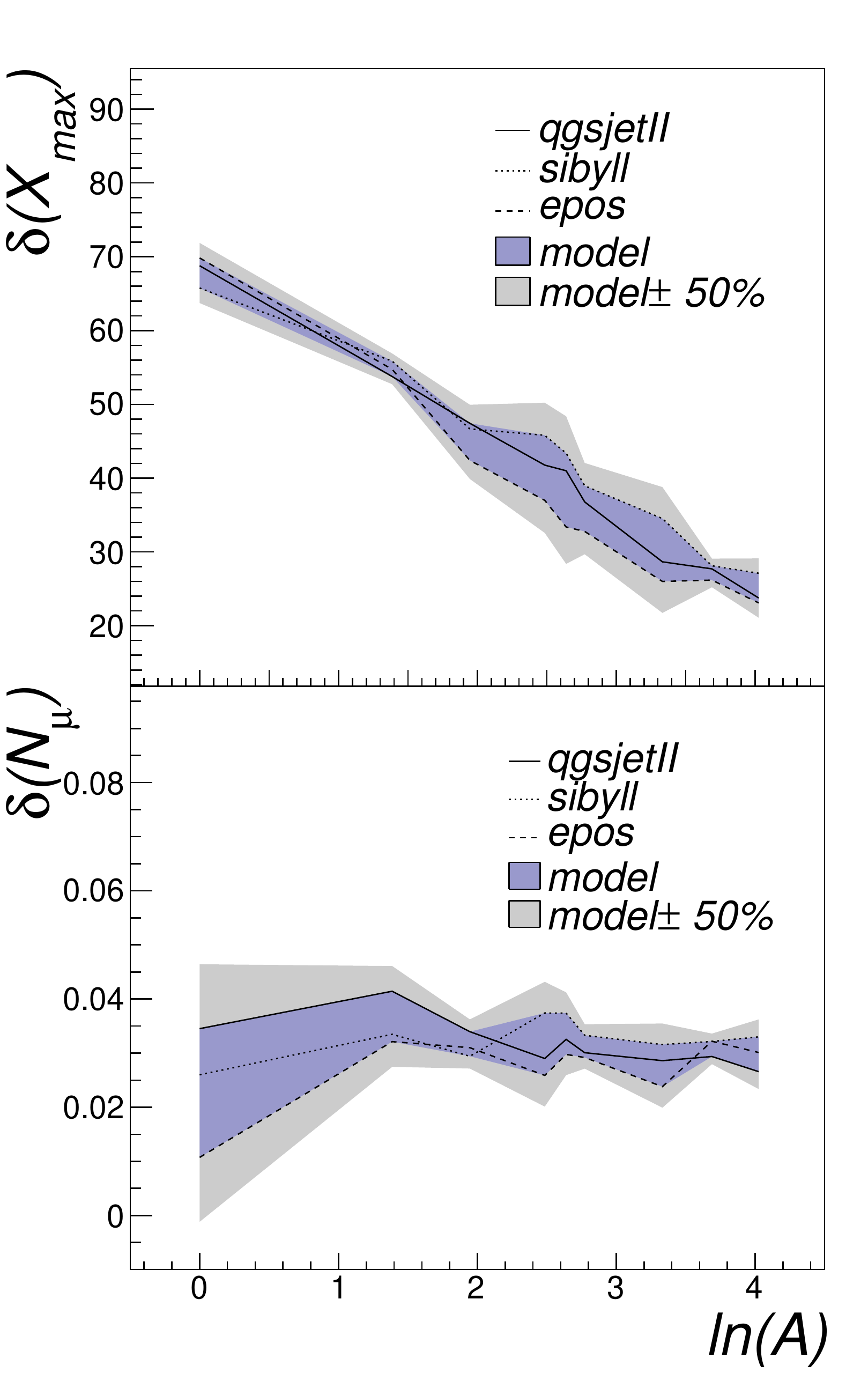}\label{DeltaConstraintFig}}
\hspace{-0.4cm}
\subtable[$\nu$ constraints]{\includegraphics[scale=0.18]{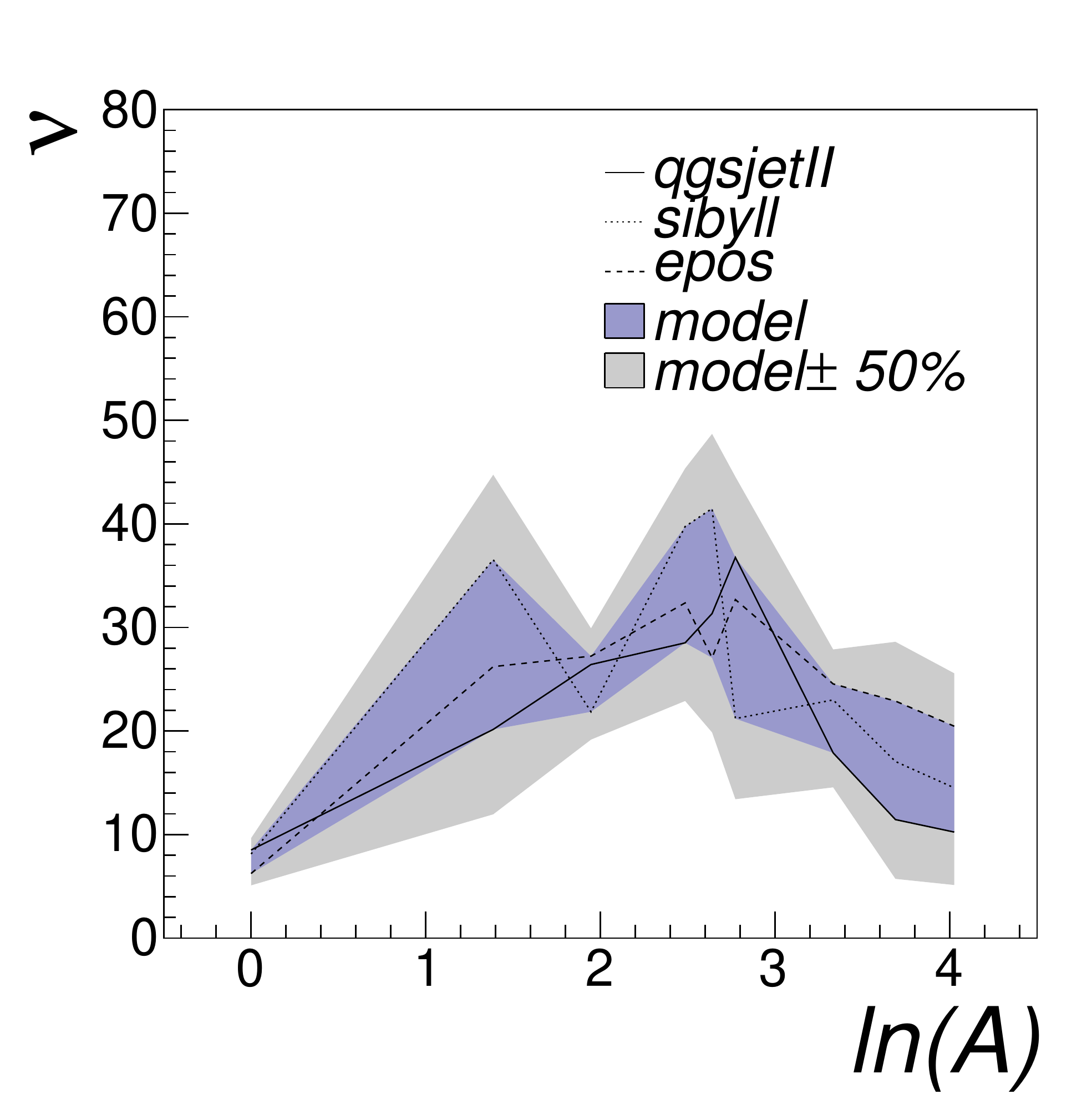}\label{NuConstraintFig}}
\vspace{-0.1cm}
\caption{Constraint region on the MST parameters from hadronic models as a function of the logarithm of the primary mass $A$.}%
\label{ParameterConstraintFig}
\end{figure}

\subsection{Model validation}\label{DataSection}
The model presented in the previous section has been validated with shower simulations generated with 
\textsc{Conex} v2r2.3 tool \cite{Conex1,Conex2} for three different hadronic models (\textsc{QgsjetII} \cite{QGSJETIIPaper}, \textsc{Sibyll} 2.1 \cite{SibyllPaper}, 
\textsc{Epos} 1.99 \cite{EPOSPaper}) 
and a large set of primary nuclei ($p$, $He$, $C$, $N$, $O$, $Si$, $Ca$, $Fe$). A fixed primary energy was assumed in the simulation, 
ranging from $10^{17}$ to 10$^{20}$ eV in step of $\log_{10}E$=0.5, and a zenith range from 0$^{\circ}$ to 90$^{\circ}$.
We restrict the analysis to the shower depth of maximum $X_\mathrm{max}$ and the number of muons $N_{\mu}$ at ground level above 1 GeV
which represent the most sensitive observables to the primary composition.\footnote{The $N_{\mu}$ observable has been scaled by an empirical parametrization $g(\theta)=p_{0}+p_{1}\theta+p_{2}\theta^{2}$ to 
get rid of the zenith angle dependence.}\\  
In Fig. \ref{MSTFitFig} we report the distribution of the two parameters obtained from \textsc{Epos} simulations for different primary nuclei 
at an energy of $10^{19}$ eV.
The agreement with the MST model is remarkable. Similar plots are obtained for the other two hadronic models (not shown here). 
In Fig. \ref{ParameterConstraintFig} the fitted values of the MST parameters are reported for the three
hadronic interaction models as a function of the logarithm of the nuclear mass. 
Although $\bmu$ and $\bSigma$ do not directly represent the mean and covariance of the distribution, it is noteworthy to retrieve the expected behavior 
of the means, decreasing with the mass for $X_\mathrm{max}$ and increasing for $N_{\mu}$, and of the variances, decreasing with the nuclear mass.
Interestingly, the skewness is found to decrease with the nuclear mass for the depth of maximum variable while assumes nearly constant values for the muon component.

\section{A mixture model for mass composition analysis}\label{ClusteringAlgorithmSection}
In this section we describe the mixture model we designed for mass composition analysis. It has been developed in \textsc{C}++
with links to the \textsc{Root} tool \cite{ROOT} for numerical integration routines and to the $\textsc{R}$ 
statistical tool \cite{RProject} via the \textsc{RInside}/\textsc{Rcpp} interface \cite{RInside} for multivariate mathematical functions. 

Assume we are provided with  a set of $N$ cosmic ray data observations $\sbX=(\bx_1, \ldots,\bx_N)$ coming from different nuclear species, 
where $\bx_i=(X_\mathrm{max_{i}}, N_{\mu_{\mathrm{i}}})$. 
In this case, the distribution of the random vector $\bX$ can be modeled via a mixture $f(\bx; \btheta)$ of $K$ multivariate
skew-$t$ distributions:
\begin{equation}
f(\bx; \btheta)= \sum_{k=1}^{K}\pi_{k}f(\bx; \bphi_k) \label{mixtureMST}
\end{equation}
where $\pi_{k}$ are the mixture  weights (satisfying $\pi_k >0$ and $\sum_{k=1}^{K}\pi_k=1$, composition abundances), $f(\bx; \bphi_k)$ denotes the
density of the $k$ component \eqref{MSTDefinition} where 
$\bphi_k=(\bmu_k, \bSigma_k, \bdelta_k, \nu_k)$ denotes the parameters of the $k$th component and $\btheta=\{\bphi_1, \ldots, \bphi_k,  \pi_1, \ldots,
\pi_k \}$ denotes the overall parameter of the mixture.

According to the maximum likelihood approach, the  parameters $\btheta$ in \eqref{mixtureMST}
can be estimated by maximizing the log-likelihood function:
\begin{equation}
\log\cL(\btheta; \sbX)= \log \prod_{i=1}^{N}f(\bx_i;\theta) = \sum_{i=1}^{N} \log f(\bx_i;\theta) \, . \label{loglik}
\end{equation}
The maximization has to be carried out numerically as no analytical solution exists. However it has been shown in 
\cite{McNicholas2012} that a closed form solution can be obtained using the EM algorithm. In next section we briefly summarize the parameter estimation 
procedure.

\begin{figure}[!h]
\centering%
\includegraphics[scale=0.22]{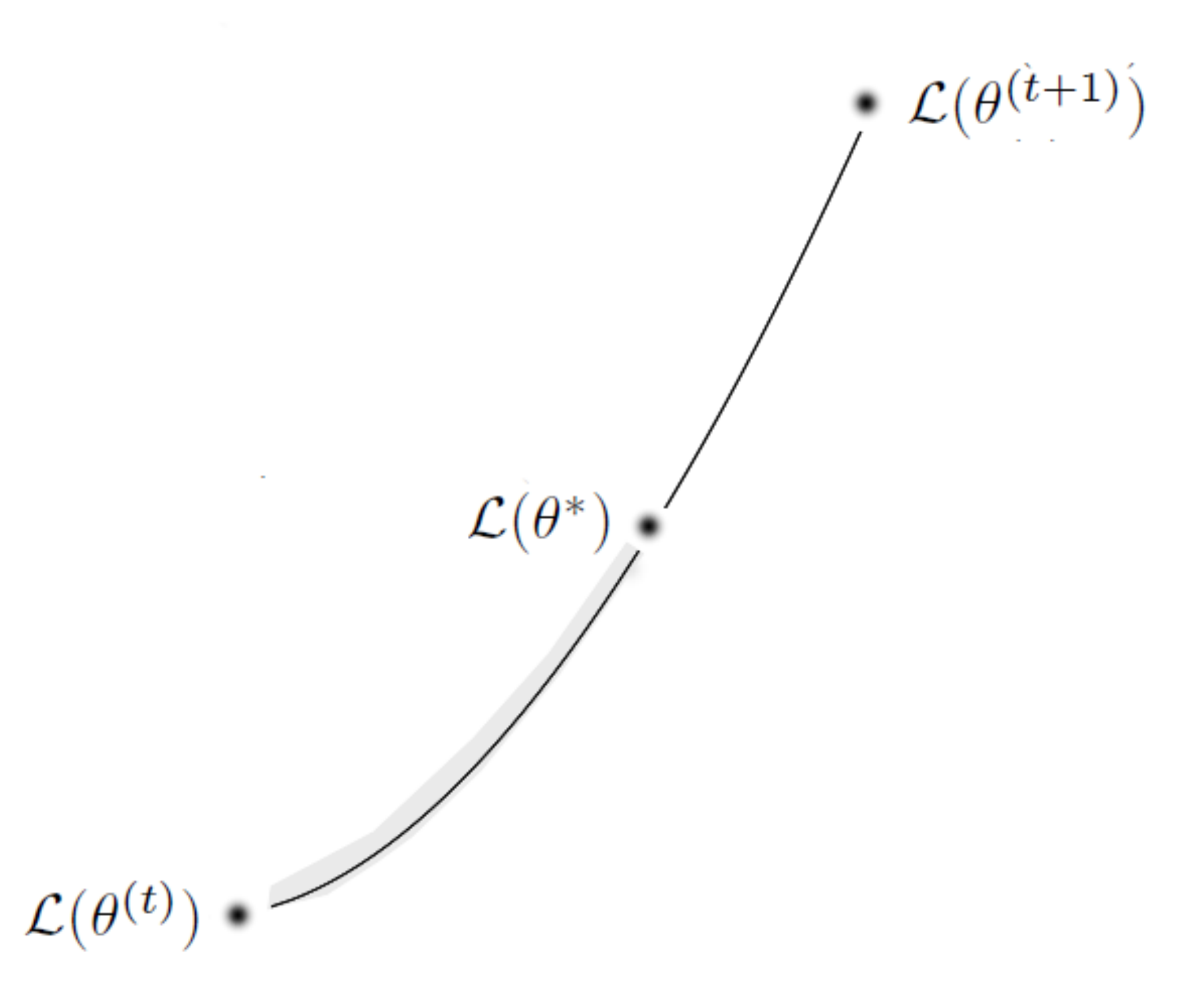}%
\vspace{-0.1cm}
\caption{Sketch showing the situation of violating constraints. At iteration (t) constraints are assumed to
be satisfied, while at the next iteration (t+1) they are violated. Given the ascent and continuous property of the 
likelihood in the EM it will exist a point (*) in between the two steps in which the violation occurs. The idea consists of retrieving such point and 
limit the EM update in the dashed gray region before.}%
\label{SketchViolatedConstraint}
\end{figure}

\subsection{EM parameter estimation}\label{MSTMixtureSection}
In the EM framework \cite{Dempster1977},  the $N$ observed data $\mathbf{x}$=($\mathbf{x}_{1}$,\dots,$\mathbf{x}_{N}$) are considered incomplete 
and $\bU=(u_1, \ldots, u_N)$ 
and $\bW=(w_1, \ldots, w_N)$ are unobservable latent variables. We introduce the missing data 
$\sbZ=(\bz_1,\ldots,\bz_N)$ such that $\bz_i=(z_{1i},\ldots,z_{Ki})$ (i=1,\dots N) with
$z_{ki}$=1 if $\bx_i$ comes from  the $k$-th component and $z_{ik}$=0 otherwise.
The set $(\sbX, \sbZ, \bU, \bW)$ denotes the complete-data. Thus  the complete-data log-likelihood $\cL_c(\btheta)$
can be expressed as:
\begin{align*}
\log\cL_c (\btheta)& = \log\cL_1(\pi_1, \ldots, \pi_K)+\log\cL_2(\bmu_1,\bSigma_1,\bdelta_1, \ldots, \bmu_K,\bSigma_K,\bdelta_K)+\log\cL_3(\nu_1, \ldots, \nu_K) 
\end{align*}
where
\begin{align*}
\log\mathcal{L}_{1}=& \sum_{k=1}^{K}\sum_{i=1}^{N}z_{ki}\log\pi_{k}\\%
\log\mathcal{L}_{2}=& \sum_{k=1}^{K}\sum_{i=1}^{N}\frac{z_{ki}}{2}[\log|\bSigma_{k}^{-1}|-\log(2\pi)+ w_{i}(\bx_{i}-\bmu_{k}-\bdelta_{k}u_{i})^{T}\bSigma_{k}^{-1}(bx_{i}-\bmu_{k}-\bdelta_{k}u_{i})]\\%
\log\mathcal{L}_{3}=& \sum_{k=1}^{K}\sum_{i=1}^{N}z_{ki}\{-\frac{1}{2}[(p-1)\log w_{i}+w_{i}u_{i}^{2}]+\\ & \quad + \frac{\nu_{k}}{2}[w_{i}-\log(\nu_{k}/2)]+\log\Gamma(\nu_{k}/2)+(\nu_{k}/2-1)\log w_{i}\}
\end{align*}
The EM algorithm is first initialized by choosing a starting approximation $\btheta^{(0)}$ for the model parameters and then proceeds 
by iterating two consecutive steps until convergence. At a given iteration $t$ the $E$ step computes the expected value 
$Q(\btheta|\btheta^{(t)})$ of the complete log-likelihood, which is maximized in the $M$ step with respect to $\btheta$ 
to obtain a new parameter estimate $\btheta^{(t+1)}$. The following expectations are required to compute $Q(\btheta|\btheta^{(t)})$:%
\begin{align*}
\mathbb{E}(Z_{ki}|\bx_{i})&= \frac{\pi_{k}^{(t)}f(\bx_{i}|\btheta_{k}^{(t)})}{\sum_{k=1}^{K}\pi_{k}^{(t)}f_{k}(\bx_{i}|\btheta_{k}^{(t)})}\equiv \tau_{ki} \\ \mathbb{E}(Z_{ki}W_{k}|\bx_{i}) &= P(Z_{ki}=1|\bx_{i})\mathbb{E}(W_{k}|x_{i}) \equiv P(Z_{ki}=1|\bx_{i})e_{1,ki}\\%
\mathbb{E}(Z_{ki}|U_{k}|W_{k}|\bx_{i})&= P(Z_{ki}=1|\bx_{i})\mathbb{E}(|U_{k}|W_{k}|\bx_{i}) \equiv P(Z_{ki}=1|\bx_{i})e_{2,ki}\\%
\mathbb{E}(Z_{ki}U_{k}^{2}W_{k}|\bx_{i})&= P(Z_{ki}=1|\bx_{i})\mathbb{E}(U_{k}^{2}W_{k}|\bx_{i}) \equiv P(Z_{ki}=1|\bx_{i})e_{3,ki}\\%
\mathbb{E}(Z_{ki}\log W_{k}|\bx_{i})&= P(Z_{ki}=1|\bx_{i})\mathbb{E}(\log W_{k}|\bx_{i}) \equiv P(Z_{ki}=1|\bx_{i})e_{4,ki}%
\end{align*}
An analytical calculation of the above expectation terms $e_{1,ki}$, $e_{2,ki}$, $e_{3,ki}$, $e_{4,ki}$ has been recently provided in 
\cite{McNicholas2012}. Maximizing $Q(\btheta|\btheta^{(t)})$, i.e. setting the derivatives with respect to $\btheta$ equal to zero, yields the
update estimates of the model parameters at the $t+1$ iteration:

\begin{align*}
\footnotesize%
\pi_{k}^{(t+1)}=& \frac{\sum_{i=1}^{N}\tau_{ki}^{(t)}}{N} \\%
\mu_{k}^{(t+1)}=& \frac{\sum_{i=1}^{N}\tau_{ki}^{(t)}(\bx_{i}e_{1,ki}^{(t)}-\delta_{k}e_{2,ki}^{(t)})}{\sum_{i=1}^{N}\tau_{ki}^{(t)}e_{1,ik}^{(t)}}\\%
\Sigma_{k}^{(t+1)}=& \frac{1}{\sum_{i=1}^{N}\tau_{ki}^{(t)}}
\{\sum_{i=1}^{N}\tau_{ki}^{(t)}[(\bx_{i}-\bmu_{k}^{(t)})(\bx_{i}-\bmu_{k}^{(t)})^{T}e_{1,ik}^{(t)}-\\%
&-e_{2,ik}^{(t)}\delta_{k}^{(t)}(\bx_{i}-\bmu_{k}^{(t)})^{T}-(\bx_{i}-\bmu_{k}^{(t)})\delta_{k}^{(t)}e_{2,ik}^{(t)}+\\%
&+e_{3,ki}^{(t)}\delta_{k}^{(t)}\delta_{k}^{(t)T}\}\\%
\delta_{k}^{(t+1)}=& \frac{\sum_{i=1}^{N}\tau_{ki}^{(t)}e_{2,ki}^{(t)}(\bx_{i}-\bmu_{k}^{(t)})}{\sum_{i=1}^{N}\tau_{ki}^{(t)}e_{3,ki}^{(t)}}\\%
(\log(\nu_{k}^{(t+1)}/2)&-\psi(\nu_{k}^{(t+1)}/2)+1)\sum_{i=1}^{N}\tau_{ki}^{(t)}+\sum_{i=1}^{N}\tau_{ki}^{(t)}(e_{4,ki}^{(t)}-e_{1,ki}^{(t)})
\end{align*}
\normalsize%
The $\nu$ update does not exist in closed form and have to be either estimated by solving numerically the above equation or specified in advance. 
The above update rules guarantee a monotonic increase of the log-likelihood.

\subsection{EM constrained}\label{ConstraintSection}

In the case of the mass 
composition identification of cosmic rays a set of parameter constraints can be derived from physical considerations as well 
as from predictions obtained with the Heitler model \cite{Heitler1954,Matthews2005} or with the current hadronic models. 
We report in Fig. \ref{ParameterConstraintFig} the mean $\mu$, covariance matrix $\Sigma$, skewness $\delta$ and degree of 
freedom $\nu$ parameters as a function of 
the logarithm of the nuclear mass $\ln(A)$ at an energy of 10$^{19}$ eV. The numerical values are obtained by fitting a single MST distribution to our 
simulated samples. The three black lines correspond to three different hadronic models, the purple area to the region constrained by the three models. 
As one can see, hadronic models provide stringent lower and upper bounds on the mixture parameters.
If we therefore specify in advance the primary masses $A_{k}$ of the mixture groups to be determined from data 
and order them according to their primary mass ($A_{k}<A_{k+1}$) we can define the following sets of constraints holding for all shower observables:

\begin{itemize}
 \item \textit{Mean constraints: $a_{\mu_{kj}} \le \mu_{kj}-\mu_{(k+1)j} \le b_{\mu_{kj}}$, $\forall k$, $j$=1,\dots,$q$} 
 where $\mu_{kj}$ denotes the $j$th component of the mean vector $\bmu_k$, for suitable constants $a_{\mu_{kj}}$, $b_{\mu_{kj}}$.\\  
 As a consequence of the higher interaction cross section of heavy nuclei with the air molecules with respect to light ones, an ordering constraint of the kind
 $\mu_{j}^{k}\lessgtr\mu_{j}^{k+1}$ is established among each nuclear component means at the same primary energy, 
 e.g. $\mu_{X_\mathrm{max}}^{k}<\mu_{X_\mathrm{max}}^{k+1}$ or $\mu_{N_{\mu}}^{k}>\mu_{N_{\mu}}^{k+1}$ and so on for other observables. 
 A more stringent constraint, requiring the difference of
 each component means within a given bound [$a_{\mu_{j}}^{k}$,$b_{\mu_{j}}^{k}$], is provided by the hadronic models.

\item \textit{Variance constraints: $\sigma_{kjj}>\sigma_{(k+1)jj}$, 
$a_{\sigma_{kjj}}\le\sigma_{kjj}\le b_{\sigma_{kjj}}$ $\forall k$, $j$=1,\dots,$q$} where $\sigma_{kjj}$ denotes the $j$th diagonal component of the 
covariance matrix $\bSigma_k$, for suitable constants $a_{\sigma_{kjj}}$,  $b_{\sigma_{kjj}}$.\\
Due to the larger number of nucleons involved in the interactions, heavy
nuclei exhibit smaller shower-to-shower fluctuations in all shower observables compared to light nuclei at the same primary energy, e.g. 
$\sigma_{X_\mathrm{max}}^{k}<\sigma_{X_\mathrm{max}}^{k+1}$ and $\sigma_{N_{\mu}}^{k}<\sigma_{N_{\mu}}^{k+1}$.
Moreover, hadronic models provides also a constraint bound [$a_{\sigma_{kj}}$,$b_{\sigma_{kj}}$] for each mixture component.

\item \textit{Skewness and degrees of freedom constraints: $a_{\delta_{kj}}\le\delta_{kj}\le b_{\delta_{kj}}$, 
$a_{\nu_{k}}\le\nu_{k}\le b_{\nu_{k}}$ $\forall k$, $j$=1,\dots,$q$} where $\delta_{kj}$ denotes the $j$th component
of the skewness vector $\bdelta_k$, with $a_{\delta_{kj}}$, $b_{\delta_{kj}}$, $a_{\nu_{kj}}$, $b_{\nu_{kj}}$
constants.\\
Useful bound regions on the skewness parameters $\delta$ and $\nu$ are provided by the hadronic models.
\end{itemize}

When analyzing simulated or real data a larger tolerance region with respect to that defined by the models should be considered for several reasons. 
First we need to account for possible discrepancies of 
the data with respect to model predictions. Moreover, if we consider the non-null experimental resolution achieved in the measurement of the shower 
observables and that the nuclear species are not considered alone in analysis but instead grouped in few general groups (i.e. light-$A$, intermediate-$A$ and heavy-$A$)  
the variance to be considered for each group $k$ is effectively larger. In Fig. \ref{ParameterConstraintFig} the tolerance region is represented by the gray area 
defined assuming a $\pm$50\% bound with respect to 
the model constraint region.
To incorporate the above constraints in the parameter space of the statistical model we consider the case of a given parameter $\mathbf{\theta}$ violating its constraints 
at a given iteration $t+1$. At the previous iteration the constraints are assumed to be satisfied. This situation is sketched in Fig. \ref{SketchViolatedConstraint}.
Due to the the continuous and monotonic 
properties of the likelihood function in the EM we can trace back the exact point $\mathbf{\theta}^{*}$ in which the constraint is violated after the $(t+1)$ update by 
introducing the following notation:
\begin{equation}
\mathbf{\theta}^{*}= (1-\alpha)\mathbf{\theta}^{(t)}+\alpha\mathbf{\theta}^{(t+1)}
\end{equation}
with $\alpha$ real values in the range [0,1]. 
When $\alpha$=0 the parameter estimated at iteration $t$ is obtained, while the update 
at iteration $t+1$ is obtained when $\alpha$=1. The constraint violation occurs at an intermediate value $\alpha$=$\alpha^{*}$. We note that when 
$\mathbf{\theta}^{(t+1)}$ 
satisfies the constraints $\mathbf{\theta}^{*}$ satisfies the constraints $\forall\alpha$.
To effectively constrain the EM update it is sufficient to choose an arbitrary $\alpha<\alpha^{*}$, e.g. $\alpha^{*}$/$s$ ($s>1$). 
In practice we are slowing down the EM algorithm with $s$ controlling the slow-down rate.
For the different types of constraints discussed above we have:
\begin{itemize}

\item \textit{Mean constraint}:\\%
$\alpha^{*}= \underset{j,k}{\min}\left\lbrace\frac{(\mu_{(k+1)j}^{(t)}-\mu_{kj}^{(t)})+a_{\mu_{kj}}}{(\mu_{(k+1)j}^{(t)}-\mu_{kj}^{(t)})-
(\mu_{(k+1)j}^{(t+1)}-\mu_{kj}^{(t+1)})}\right\rbrace$,  
$\alpha^{*}= \underset{j,k}{\min}\left\lbrace\frac{(\mu_{(k+1)j}^{(t)}-\mu_{kj}^{(t)})+b_{\mu_{kj}}}{(\mu_{(k+1)j}^{(t)}-\mu_{kj}^{(t)})-
(\mu_{(k+1)j}^{(t+1)}-\mu_{kj}^{(t+1)})}\right\rbrace$

\item \textit{Variance constraint}:\\%
$\alpha^{*}= \underset{j,k}{\min}\left\lbrace\left[1-\frac{\sigma_{(k+1)jj}^{(t+1)}-\sigma_{kjj}^{(t+1)}}{\sigma_{(k+1)jj}^{(t)}-
\sigma_{kjj}^{(t)}}\right]^{-1}\right\rbrace$\\

$\alpha^{*}= \underset{j}{\min}\left\lbrace\frac{a_{\sigma_{kjj}}-\sigma_{kjj}^{(t)}}{\sigma_{kjj}^{(t+1)}-\sigma_{kjj}^{(t)}}\right\rbrace$, 
$\alpha^{*}= \underset{j}{\min}\left\lbrace\frac{b_{\sigma_{kjj}}-\sigma_{kjj}^{(t)}}{\sigma_{kjj}^{(t+1)}-\sigma_{kjj}^{(t)}}\right\rbrace$ \;\;\;$\forall k$

\item \textit{Skewness constraint}:\\%
$\alpha^{*}= \underset{j}{\min}\left\lbrace\frac{a_{\delta_{kj}}-\delta_{kj}^{(t)}}{\delta_{kj}^{(t+1)}-\delta_{kj}^{(t)}}\right\rbrace$,
$\alpha^{*}= \underset{j}{\min}\left\lbrace\frac{b_{\delta_{kj}}-\delta_{kj}^{(t)}}{\delta_{kj}^{(t+1)}-\delta_{kj}^{(t)}}\right\rbrace$ \;\;\;$\forall k$

\item \textit{Degrees of freedom constraint}:\\%
$\alpha^{*}= \left(\frac{a_{\nu_{k}}-\nu_{k}^{(t)}}{\nu_{k}^{(t+1)}-\nu_{k}^{(t)}}\right)$,
$\alpha^{*}= \left(\frac{b_{\nu_{k}}-\nu_{k}^{(t)}}{\nu_{k}^{(t+1)}-\nu_{k}^{(t)}}\right)$ \;\;\;$\forall k$
\end{itemize}

\section{Method application to data}\label{AnalysisSection}
In this section we report the fit results (parameter estimation accuracy and classification performances) obtained over random sets ($N_{samples}$=100) of 
simulated data, generated with the \textsc{Epos} 1.99 model for a reference energy of 10$^{19}$ eV. Two kind of samples have been produced with the following 
specifications:
\begin{itemize}
 \item \textit{Set I}: 1000 two-dimensional observations of 3 nuclei ($p$, $N$, $Fe$) with relative abundances set to 0.5, 0.2, 0.3 respectively.
 \item \textit{Set II}: 1000 two-dimensional observations of 5 nuclei ($p$, $He$, $N$, $Si$, $Fe$) with relative abundances set to 0.4, 0.1, 0.1,
 0.1, 0.3 respectively. Each observation has been convoluted with a gaussian distribution of width $\sigma_{det}$ to take into account the effect of a non-zero 
 experimental resolution. We assumed realistic resolutions for the two variables, namely $\sigma_{det}(X_\mathrm{max})$= 20 g/cm$^{2}$ and a comparable resolution 
 for the number of muons $\sigma_{det}(N_{\mu})$= 3\%.
\end{itemize}
The fit of the data are repeated in three different conditions. In a first case we assume that the hadronic models are giving a reasonable representation of the data, 
hence we fixed all mixture parameters and determine the component fractions. This case, denoted as \emph{$\pi$ fit} in the following, corresponds to what has been typically done so far when analyzing real 
cosmic ray data. In a second case, denoted as \emph{$(\pi+\mu)$ fit}, we assume that the models are not ``trustable'' in the mean parameters while continue to provide a reasonable description of the data 
for what concerns the shape of the shower observable distributions. This situation presumably corresponds to what has been recently observed in real data 
for the muon number observable. We therefore fitted the means and component weights and leave the other parameters fixed. Finally in the third case, denoted as
\emph{$(\pi+\mu+\Sigma+\delta)$ fit}, we released 
all parameters in the fit but the number of degrees of freedom which is specified in advance.\\

\begin{figure}[!t]
\centering
\subtable[Data Set I]{\includegraphics[scale=0.35]{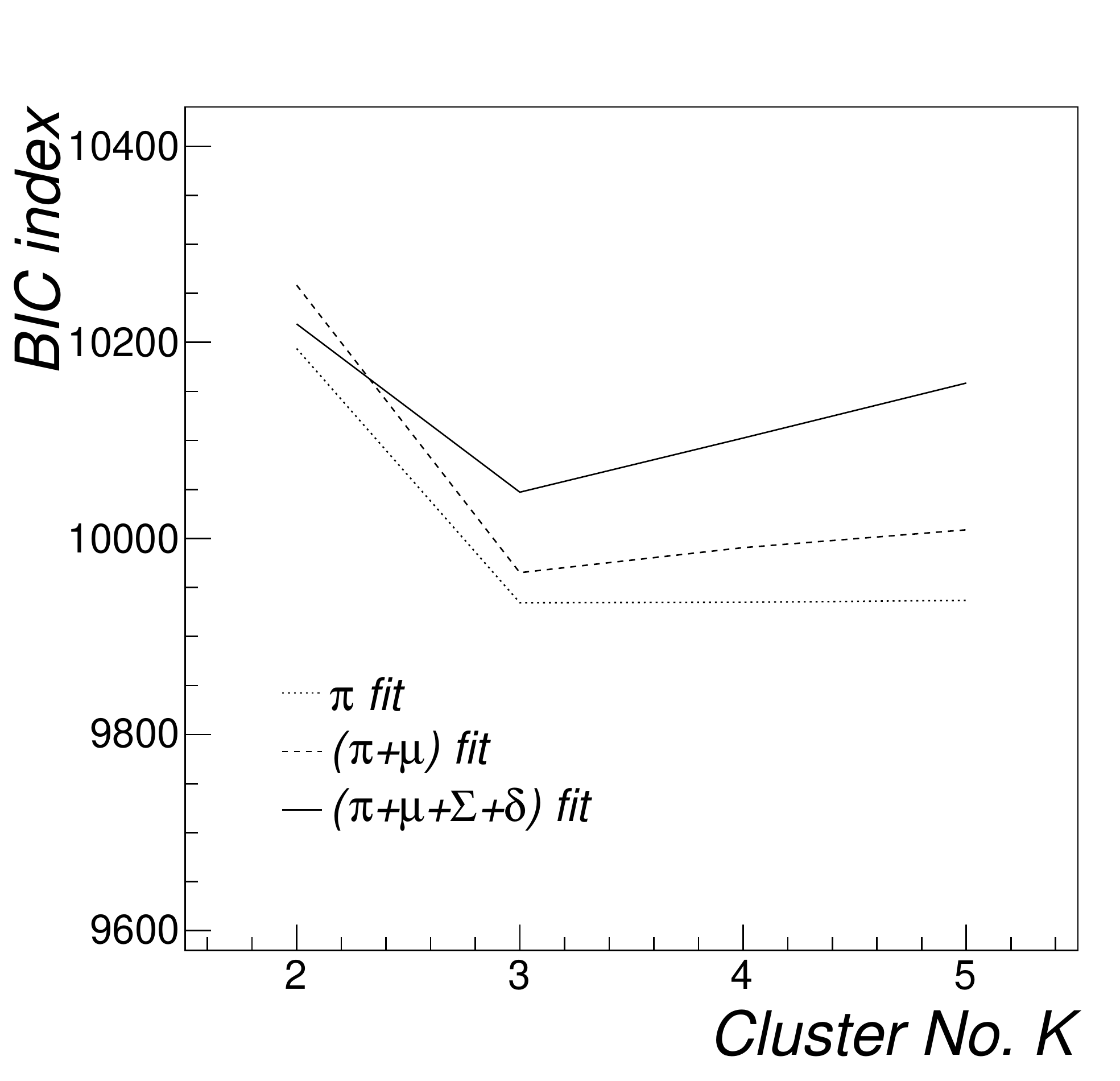}\label{BICIndexFig1}}
\hspace{-0.4cm}
\subtable[Data Set II]{\includegraphics[scale=0.35]{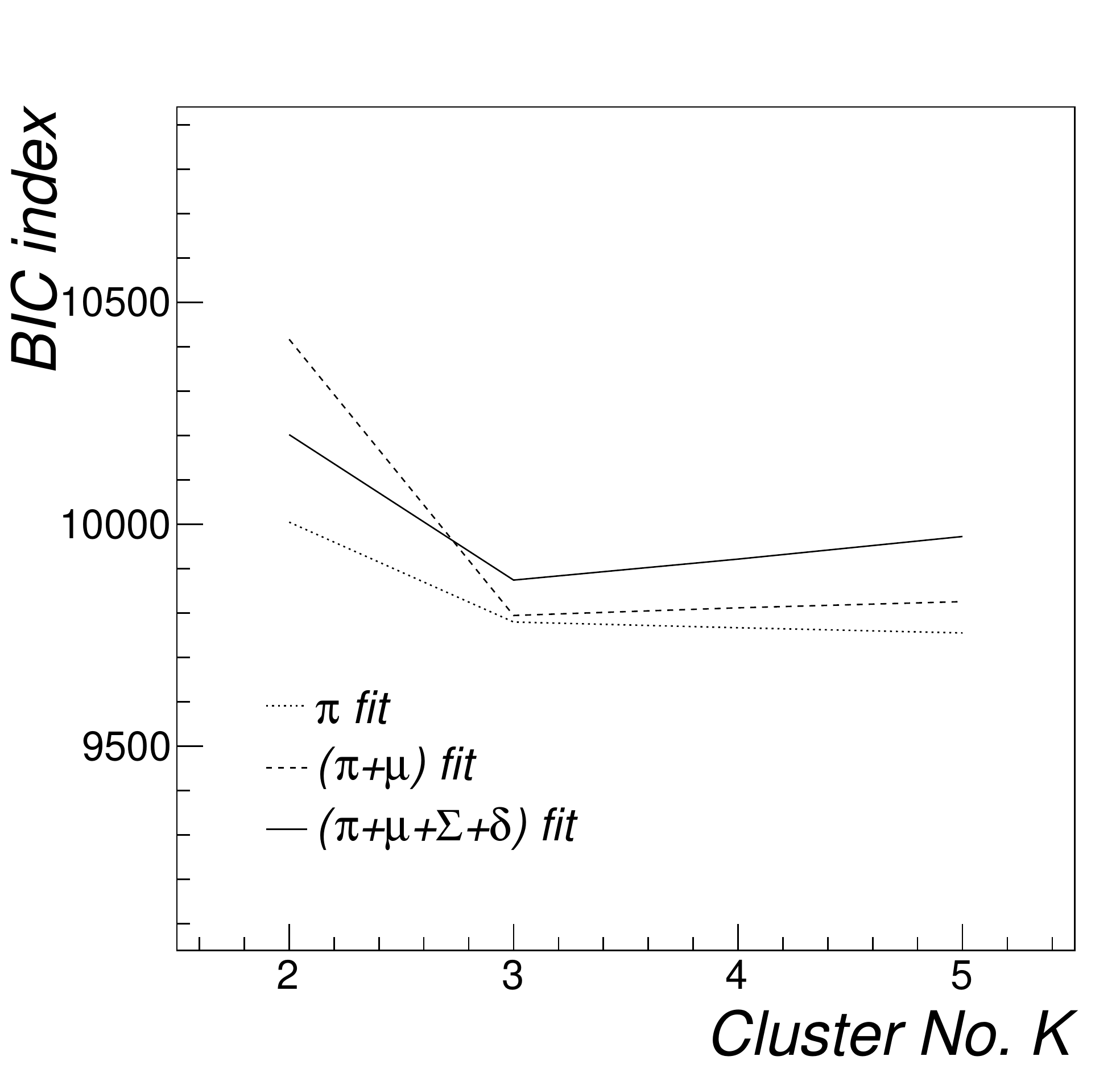}\label{BICIndexFig2}}
\vspace{-0.1cm}
\caption{BIC index computed for data sample I (left panel) and II (right panel) as a function of the number of components $K$ for the three fitting
assumptions: $\pi$ fit (dotted lines), $\pi$+$\mu$ (dashed lines), $\pi$+$\mu$+$\Sigma$+$\delta$ fit (solid lines).}%
\label{BICIndexFig}
\end{figure}

The parameters are initialized with multiple random starting values ($\sim$100) generated 
in the constrained space. The best starting values in terms of the maximum log-likelihood achieved are then used for the final fit estimate.\\
Each optimization is considered to have converged if at a given iteration stage it satisfies the Aitken criterion \cite{Aitken}, namely when 
$\mathcal{L}_{\infty}^{(t+1)}$-$\mathcal{L}^{(t+1)}<\epsilon$ ($\epsilon$=10$^{-6}$), where $\mathcal{L}^{(t+1)}$ is the log-likelihood at iteration $t+1$ 
and $\mathcal{L}_{\infty}^{(t+1)}$ is the asymptotic log-likelihood at iteration $t+1$ \cite{Bohning}:
\begin{equation}
\mathcal{L}_{\infty}^{(t+1)}=  \mathcal{L}^{(t)} + \frac{\mathcal{L}^{(t+1)}-\mathcal{L}^{(t)}}{1-a^{(t)}}\;\;\;\;
\end{equation}
with $a^{(t)}= (\mathcal{L}^{(t+1)}-\mathcal{L}^{(t)})/(\mathcal{L}^{(t)}-\mathcal{L}^{(t-1)})$ Aitken acceleration at iteration $t$.

\begin{figure}[!th]
\centering%
\subtable[Fit - Data Set I]{\includegraphics[scale=0.28]{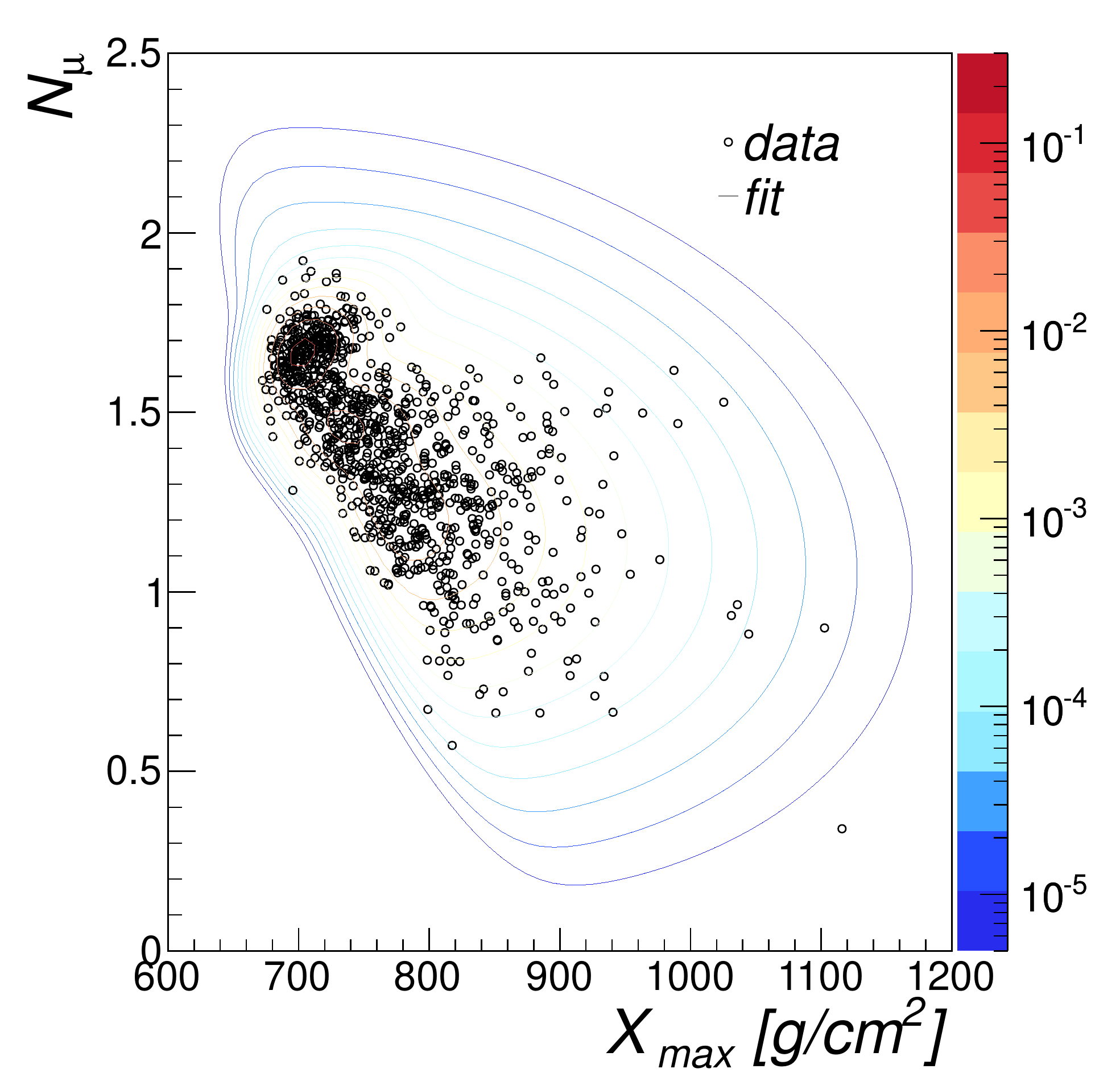}\label{FitResultsFig1}}%
\subtable[Fit - Data Set II]{\includegraphics[scale=0.28]{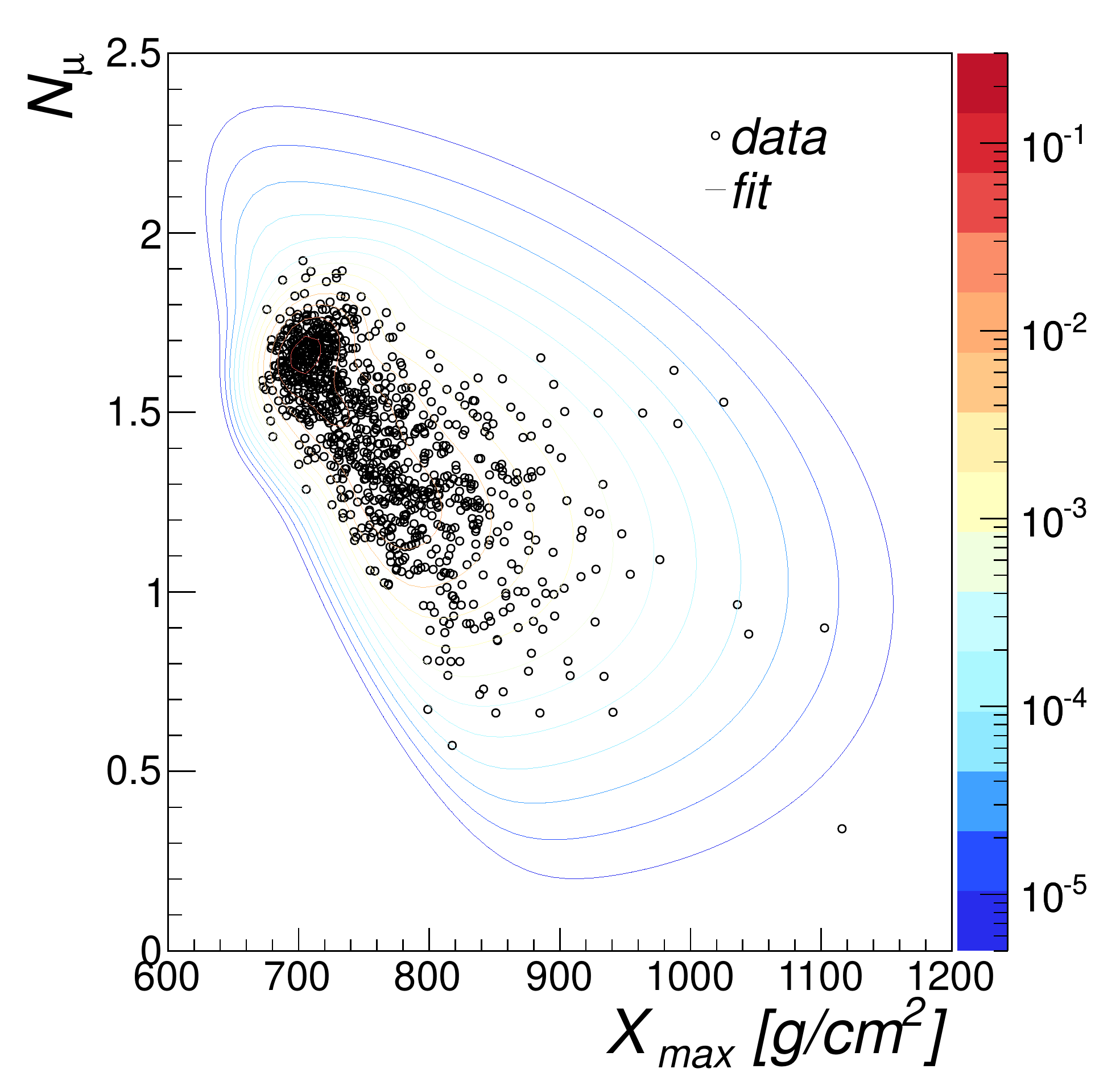}\label{FitResultsFig2}}\\%
\subtable[Fit components - Data Set I]{\includegraphics[scale=0.43]{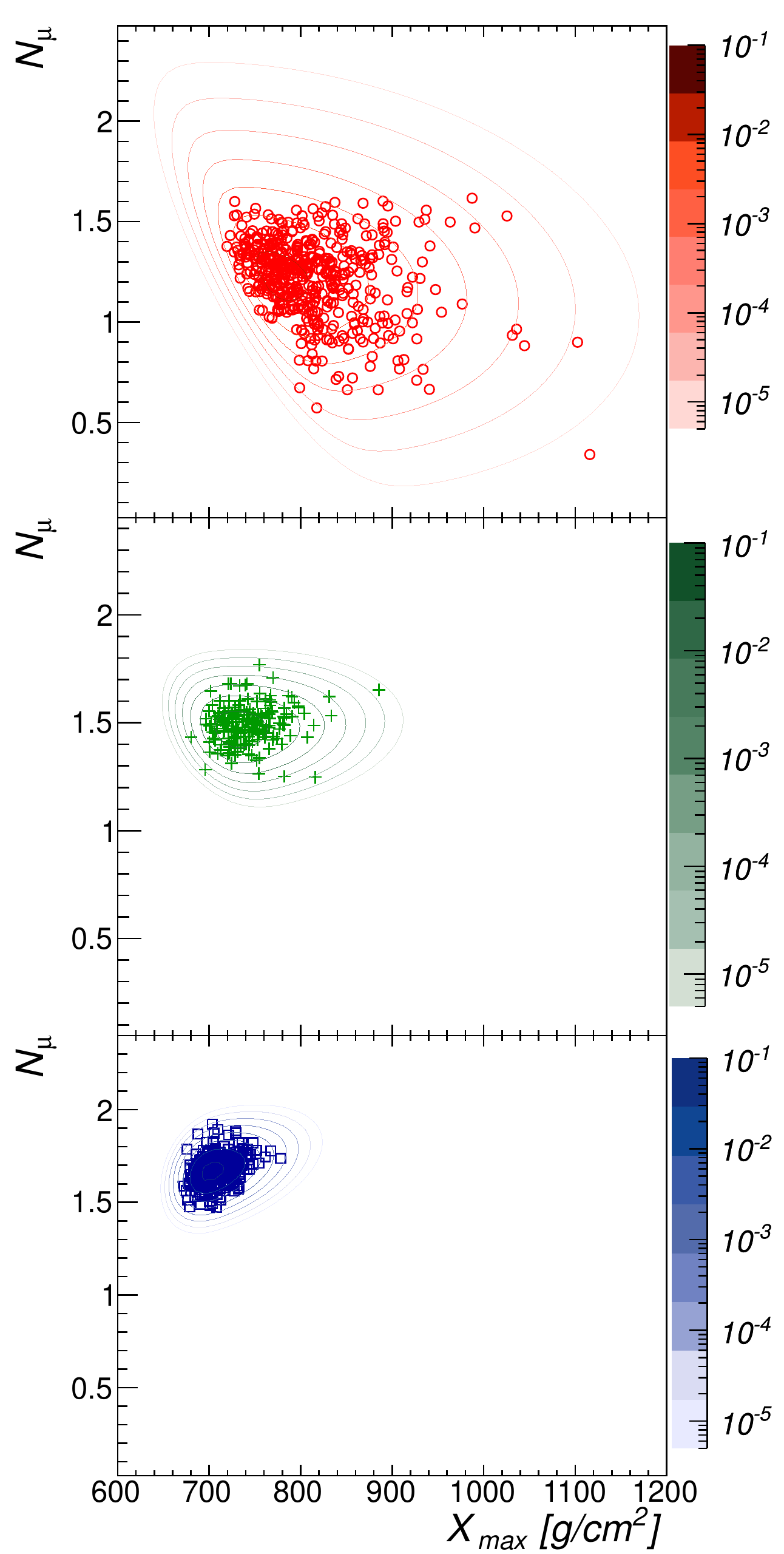}\label{FitComponentsResultsFig1}}%
\subtable[Fit components - Data Set II]{\includegraphics[scale=0.43]{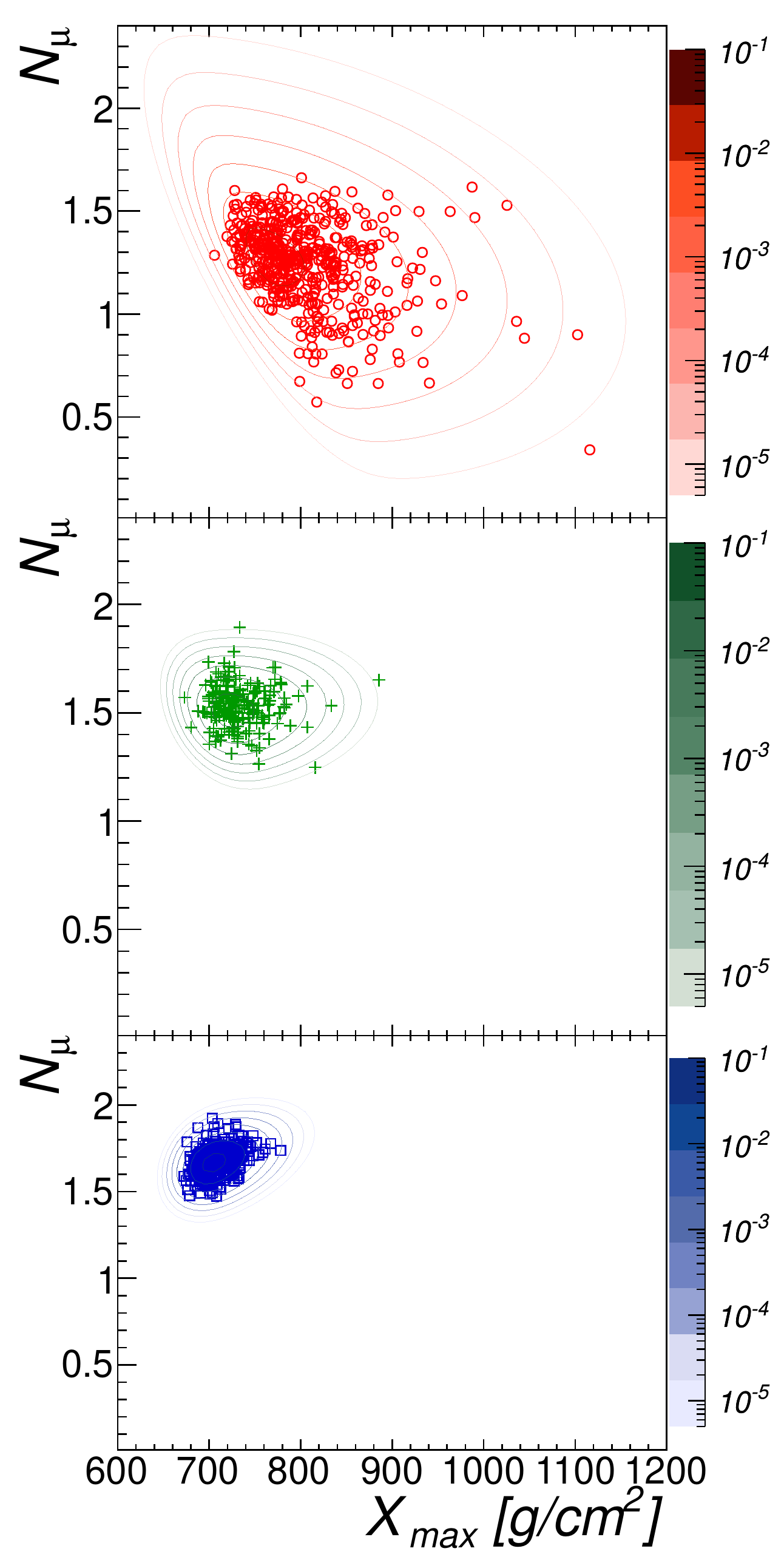}\label{FitComponentsResultsFig2}} %
\vspace{-0.1cm}
\caption{Fit results for a data sample generated from set I (left panel) and II (right panel) for the full fit ($\pi$+$\mu$+$\Sigma$+$\delta$) assumption. 
Top panels: Contour plots of the 3-component MST mixture model fitted to data (empty black dots);
Bottom panels: Contour plots of each fitted MST component (red=$p$, green=$N$, blue=$Fe$). Colored dots indicate the true observations from each mass group (see text).}%
\label{FitResultsFig}
\end{figure}

\subsection{Choice of the number of components}  
Several criteria are available in literature to estimate the number of clusters in the data. However, none of these generally leads to definite results, 
i.e. on the same data set different criteria can predict a different number of optimal clusters, and a user decision
is anyway required according to the particular requirements of the analysis. In our case the goal is to provide
a best fit of the data with the minimum number of components and the best performances, in terms of classification and mass abundance reconstruction, 
at least for the light-heavy groups or eventually for three category groups (light-intermediate-heavy).\\
It is however instructive to test at least one of the available criteria on the market to our problem, for example the Bayesian Information Criterion (\textsc{bic}).
The \textsc{bic} index is defined as 
the maximized value of the likelihood $\log\hat{\mathcal{L}}$ plus a penalty term accounting for possible overfitting of the data 
as the number of components (model parameters) increases:
\begin{equation}
\mbox{\textsc{bic}}= -2\log\hat{\mathcal{L}} + k\log N
\end{equation}
where $k$ the number of degrees of freedom of the model, e.g. the number of free parameters, and $N$ the number of observations. According
to this definition, the model with the larger \textsc{bic} value is the one to be preferred.
We report in Fig. \ref{BICIndexFig} the BIC index computed for data sample I (left panel) and II (right panel) for the three fitting assumptions, respectively reported
with dotted, dashed and solid lines. As one can see, the criterion effectively
manages to predict the real number of components present in the first data sample. Further, it clearly evidences that both data samples cannot 
be efficiently described with a 2-component model and that 3 components are sufficient to provide an accurate description of data generated by a larger 
number of components, as in the second data set. We will therefore report in next section the fitting results relative to a 3 component fit.

\begin{table}[!th]
\centering%
\footnotesize%

\begin{tabular}{c|c c c|c c c|c c c|}
\hline%
& \multicolumn{3}{>{}c|}{\textbf{$\mathbf{\pi}$ Fit}} %
& \multicolumn{3}{>{}c|}{\textbf{($\mathbf{\pi}$+$\mathbf{\mu}$) Fit}}%
& \multicolumn{3}{>{}c|}{\textbf{($\mathbf{\pi}$+$\mathbf{\mu}$+$\mathbf{\Sigma}$+$\mathbf{\delta}$) Fit}}\\%
\textit{Par} & \multicolumn{3}{>{}c|}{\textit{Component}} & \multicolumn{3}{>{}c|}{\textit{Component}} & \multicolumn{3}{>{}c|}{\textit{Component}}\\%
\hline%
$\pi$ & 0.51 & 0.19 & 0.30 &  0.50 & 0.20 & 0.30 & 0.50 & 0.22 & 0.28 \\ %
$\mu_{1}$ & 745.86 &  716.13 & 689.51 & 751.11 & 715.39 & 689.71 & 751.38 & 709.23 & 688.96\\%
$\mu_{2}$ & 1.23 & 1.47 & 1.65 &  1.23 & 1.45 & 1.64 & 1.22 & 1.46 & 1.65\\%
$\sigma_{11}$ & 310.56 & 175.88 & 90.19 &  310.56 & 175.88 & 90.19 & 518.71 & 237.59 & 84.18\\%
$\sigma_{22}$ & 0.027 &  0.0066 & 0.0047 & 0.027 &  0.0066 & 0.0047 &
0.032 & 0.0064 & 0.0044\\%
$\sigma_{12}$ & -2.19 & 0.03 & 0.08 & -2.19 & 0.03 & 0.08 & -3.47 & -0.52 & 0.17\\%
$\delta_{1}$ & 69.84 & 33.38 & 23.09 & 69.84 & 33.38 & 23.09 & 66.15 & 41.78 & 24.42\\%
$\delta_{2}$ & 0.01 & 0.03 & 0.03 & 0.01 & 0.03 & 0.03 & 0.02 & 0.03 & 0.03\\%
$\nu$ & 6.23 & 27.07 & 20.46 & 6.23 & 27.07 & 20.46 & 6.23 & 27.07 & 20.46\\%
$\mathcal{L}$ & \multicolumn{3}{>{}c|}{-4964.39} &  \multicolumn{3}{>{}c|}{-4487.93} & \multicolumn{3}{>{}c|}{-4947.89}\\%
$\varepsilon_{tot}$ & \multicolumn{3}{>{}c|}{0.92} & \multicolumn{3}{>{}c|}{0.91} & \multicolumn{3}{>{}c|}{0.91}\\%
$\varepsilon$ & 0.93 & 0.82 & 0.97  & 0.91 & 0.81 & 0.97 & 0.90 & 0.87 & 0.94\\%
\hline%
\end{tabular}
\caption{Fit results for data set I}
\label{DataSetIFitTable}
\end{table}

\begin{table}[!th]
\centering%
\footnotesize%

\begin{tabular}{c|c c c|c c c|c c c|}
\hline%
& \multicolumn{3}{>{}c|}{\textbf{$\mathbf{\pi}$ Fit}} %
& \multicolumn{3}{>{}c|}{\textbf{($\mathbf{\pi}$+$\mathbf{\mu}$) Fit}}%
& \multicolumn{3}{>{}c|}{\textbf{($\mathbf{\pi}$+$\mathbf{\mu}$+$\mathbf{\Sigma}$+$\mathbf{\delta}$) Fit}}\\%
\textit{Par} & \multicolumn{3}{>{}c|}{\textit{Component}} & \multicolumn{3}{>{}c|}{\textit{Component}} & \multicolumn{3}{>{}c|}{\textit{Component}}\\%
\hline%
$\pi$ & 0.47 & 0.20 & 0.34  &   0.49 & 0.18 & 0.33    &   0.53 & 0.18 & 0.29 \\%
$\mu_{1}$ & 745.86 & 716.13 & 689.51 & 743.79 & 710.56 & 690.34 & 741.39 & 703.41 & 691.91\\%
$\mu_{2}$ & 1.23 & 1.47 & 1.65 & 1.25 & 1.48 & 1.64 & 1.28 & 1.50 & 1.65\\%
$\sigma_{11}$ & 310.56 & 175.88 & 90.19 & 310.56 & 175.88 & 90.19 & 518.71 & 209.70 & 113.42\\%
$\sigma_{22}$ & 0.027 & 0.0066 & 0.0047 & 0.027 & 0.0066 & 0.0047 & 0.033 & 0.0066 & 0.0046\\%
$\sigma_{12}$ & -2.19 & 0.03 & 0.08 & -2.19 & 0.03 & 0.08 & -3.43 & -0.45 & 0.17\\%
$\delta_{1}$ & 69.84 & 33.38 & 23.09 & 69.84 & 33.38 & 23.09 & 64.27 & 37.27 & 21.08\\%
$\delta_{2}$ & 0.01 & 0.03 & 0.03 & 0.01 & 0.03 & 0.03 & -0.001 & 0.03 & 0.03\\%
$\nu$ & 6.23 & 27.07 & 20.46 & 6.23  & 27.07 & 20.46 & 6.23  & 27.07 & 20.46\\%
$\mathcal{L}$ & \multicolumn{3}{>{}c|}{-4886.3} &  \multicolumn{3}{>{}c|}{-4881.5} & \multicolumn{3}{>{}c|}{-4870.27}\\%
$\varepsilon_{tot}$ & \multicolumn{3}{>{}c|}{0.86} & \multicolumn{3}{>{}c|}{0.88} & \multicolumn{3}{>{}c|}{0.87}\\%
$\varepsilon$ & 0.87 & 0.66 & 0.97 & 0.92 & 0.65 & 0.97 & 0.94 & 0.64 & 0.93\\%
\hline%
\end{tabular}
\caption{Fit results for data set II}
\label{DataSetIIFitTable}
\end{table}

\subsection{Fitting performances}  
\begin{figure}[!t]
\centering%
\subtable[Composition Fit - Data Set I]{\includegraphics[scale=0.35]{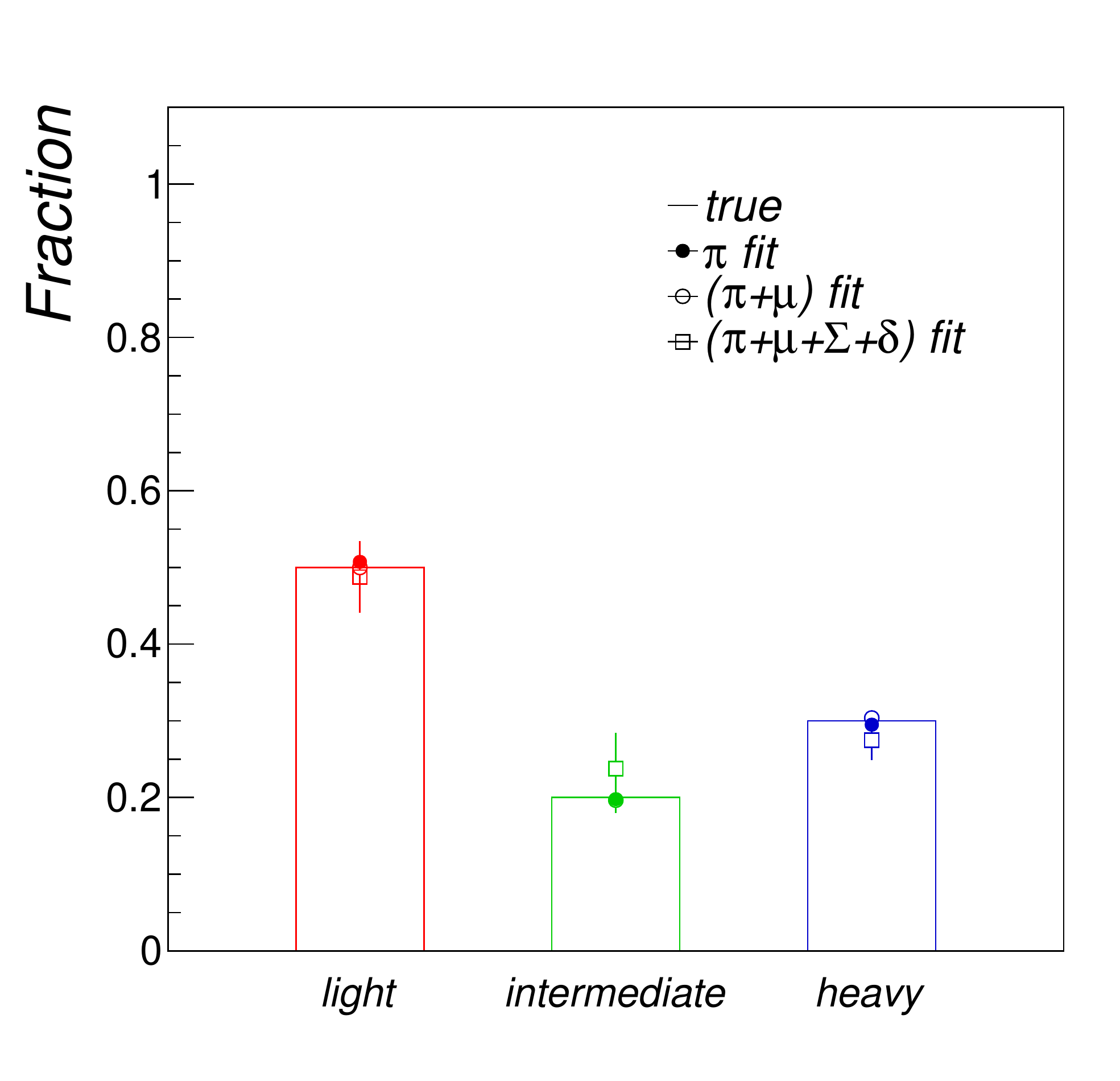}\label{FractionFitPerformanceFig1}}%
\hspace{-0.2cm}
\subtable[Classification - Data Set I]{\includegraphics[scale=0.35]{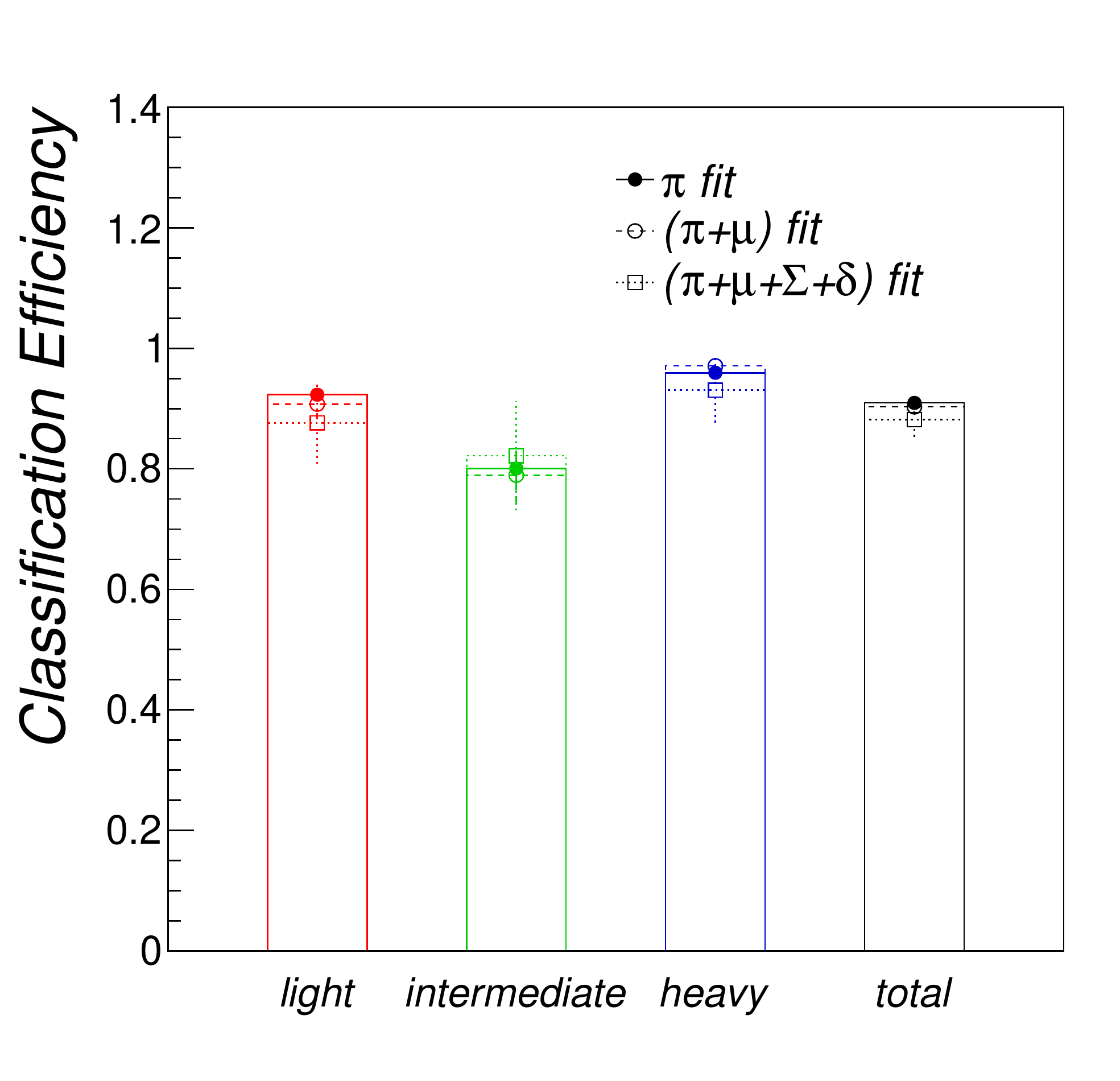}\label{ClassificationPerformanceFig1}}\\%
\vspace{-0.4cm}
\subtable[Composition Fit - Data Set II]{\includegraphics[scale=0.35]{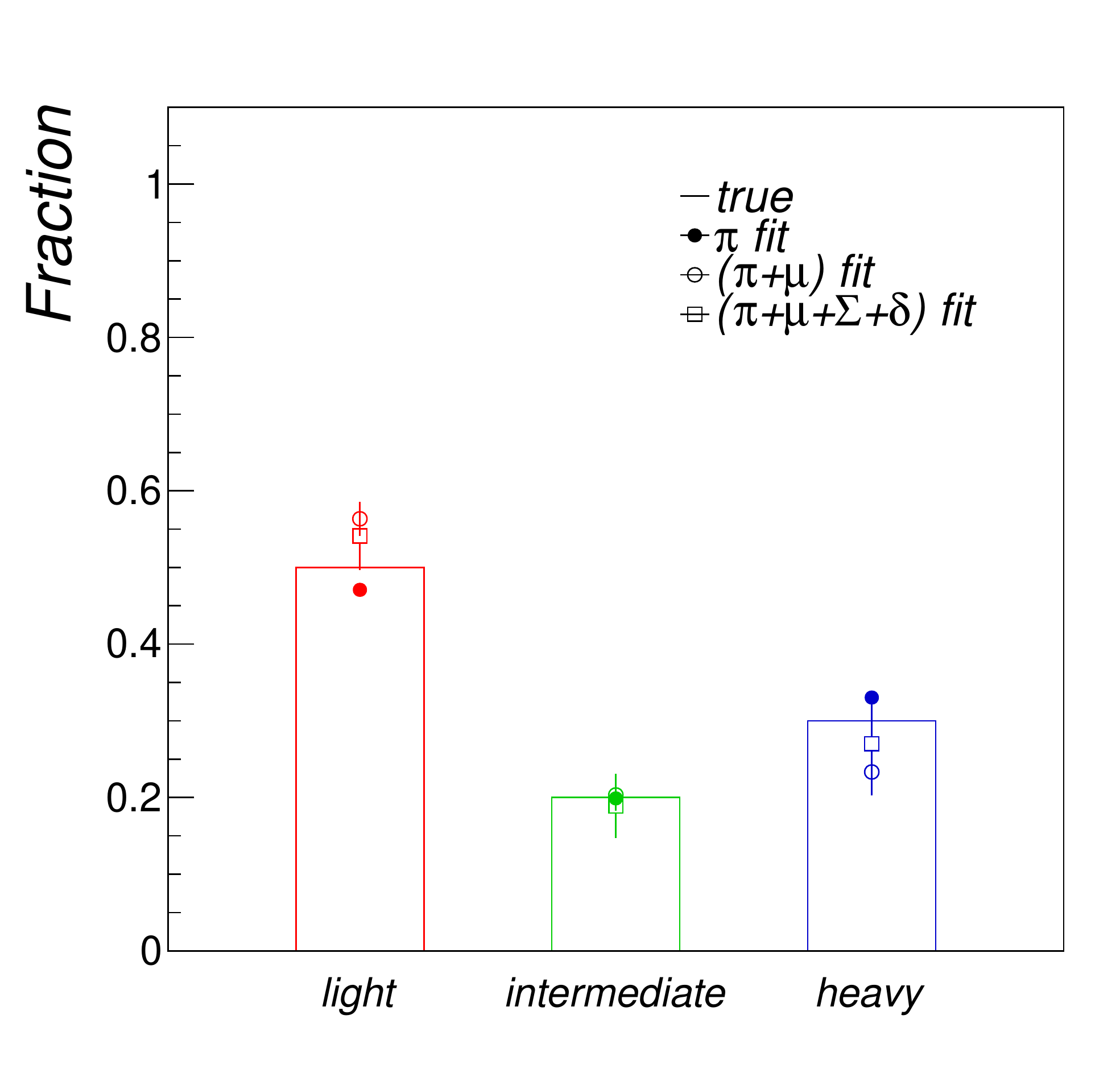}\label{FractionFitPerformanceFig2}}%
\hspace{-0.2cm}%
\subtable[Classification Fit - Data Set II]{\includegraphics[scale=0.35]{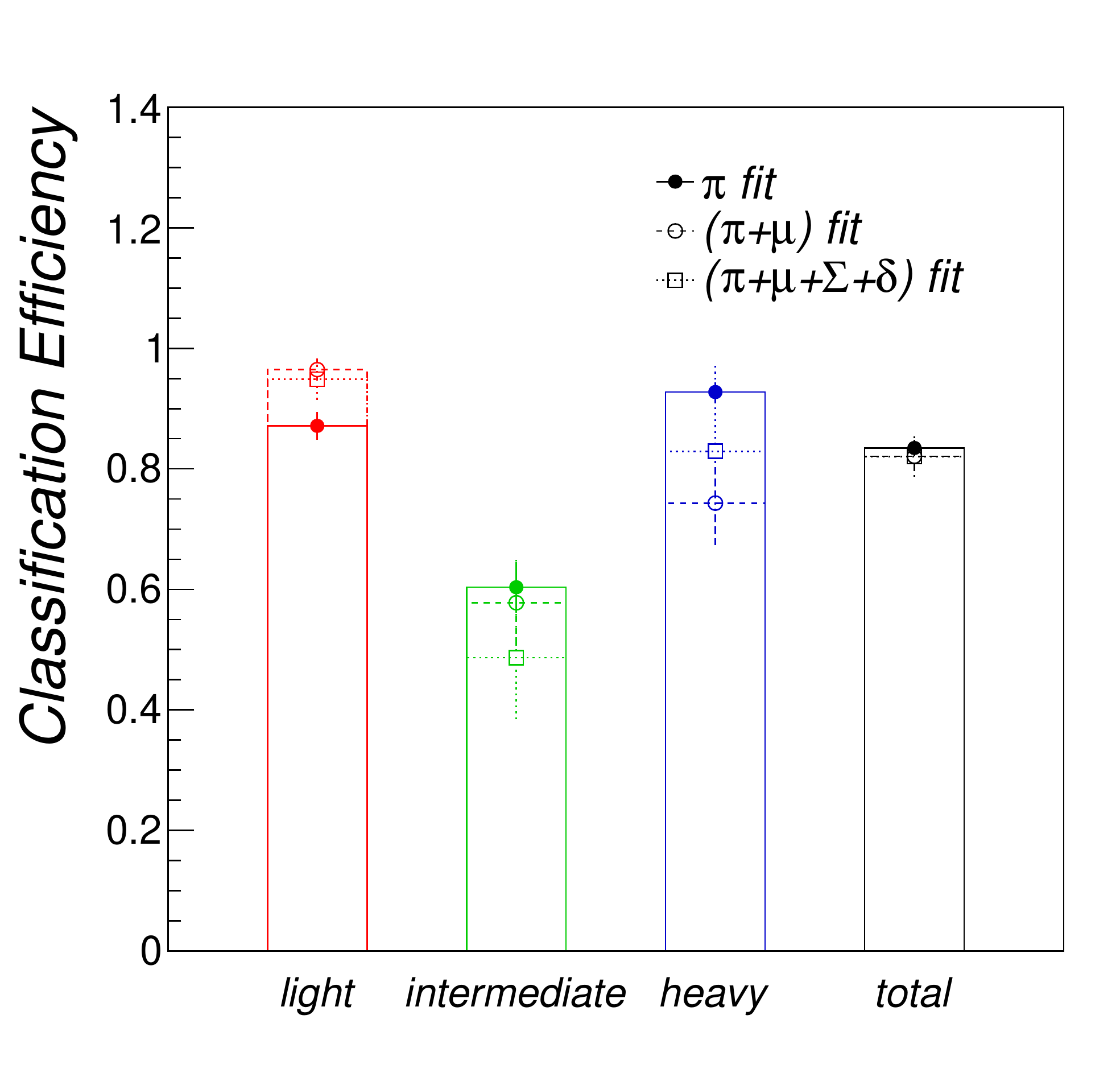}\label{ClassificationPerformanceFig2}} %
\vspace{-0.1cm}
\caption{Performance of composition abundance reconstruction and classification for data set I (upper panels) and II (bottom panels) for the three
fitting conditions ($\pi$ fit: filled dots, ($\pi$+$\mu$) fit: empty dots, ($\pi$+$\mu$+$\Sigma$+$\delta$) fit: empty squares) and 3-component groups 
(light: red, intermediate: green, heavy: blue).}%
\label{FitPerformanceFig}
\end{figure}

In Fig. \ref{FitResultsFig} we report the results obtained by fitting a 3-components MST mixture model ($p$+$N$+$Fe$) to one particular sample generated 
from data set I (left panels) and data set II (right panels). For simplicity we present the results relative to the full fit situation, as 
the fit results obtained with the other two assumptions are visually similar, and report the values of the fitted parameters with the three fitting assumptions
in Tables \ref{DataSetIFitTable} and \ref{DataSetIIFitTable}. The colored lines represent the contour plots of the fitted model. 
We consider 3 general groups to be fitted: light (1$\le$A$\le$4), intermediate (12$\le$A$\le$28) and heavy (28$<$A$\le$56). 
Lower panels show separately the three groups present in the data with the fitted components superimposed (light: red lines, intermediate: green, heavy: blue).
As can be seen in both cases the fit nicely converges towards the expected solution defined by fitting a single MST model to each component alone. 
\\To evaluate the performances achieved for composition reconstruction we report in Figures \ref{FractionFitPerformanceFig1} and 
\ref{FractionFitPerformanceFig2} the values of the fitted fractions averaged over the $N_{samples}$ generated samples for three mass groups and
fitting conditions ($\pi$ fit: filled dots, $\pi$+$\mu$ fit: empty dots, $\pi$+$\mu$+$\Sigma$+$\delta$ fit: empty squares). The histograms shown with solid lines
indicate the true composition fractions. The error bars refer to the obtained fraction RMS. For both data sets the method is able to resolve 
with good accuracy the three mass groups, with a slightly larger deviation for data set II with respect to the expected values, due to the helium and silicon 
contamination. Such small bias is however within the uncertainties of the method, found of the order of 0.05 on the reconstructed fractions.

\subsection{Classification performances}   
The fitted model can be used also for classification scopes. Each observation $i$ is assigned to the mixture component $k$
according to the maximum a posteriori probability $\tau_{ik}$. We can therefore compute for both data sets the efficiency $\varepsilon$ 
achieved for classification. We consider here, as above, the same 3 groups to be identified: light (1$\le$A$\le$4), intermediate (12$\le$A$\le$28) 
and heavy (28$<$A$\le$56). In Tables \ref{DataSetIFitTable} and \ref{DataSetIIFitTable} we reported the results for each mixture component separately in the case of the data sample 
analyzed in previous paragraph.\\The average classification performances achieved in both data sets are reported in Figures \ref{ClassificationPerformanceFig1} and
\ref{ClassificationPerformanceFig2} for the three groups (light: red, intermediate: green, heavy: blue).
In data set I an overall classification efficiency $\varepsilon\sim$0.9 is observed for all fitting cases with a slightly larger error rate for the intermediate 
component ($\sim$20\%). In data set II we have a considerably larger contamination coming from other species with respect to that used to fit the data and hence 
the event-by-event classification performances are significantly deteriorated for the intermediate component, with typical error rates of $\sim$40\%. 
However, focusing on the two extreme mass groups to be determined, we still manage to achieve a good classification with efficiencies around 90\%.
\\The obtained misclassification errors are comparable with typical results obtained with completely supervised methods, such as neural networks, 
in \cite{RiggiNIMPreparation}.

\section{Summary}\label{SummarySection}
In the present paper we proposed a new model, based on the multivariate skew-$t$ function, to describe the joint distribution of cosmic ray shower 
observables at high energies. We have tested our model with simulated data for a large set of nuclei, 
focusing the attention on the depth of shower maximum and number of muons variables, which are the most discriminating in composition studies. 
We also developed a constrained clustering algorithm to reconstruct the mass composition information
of the data, on an event-by-event basis as well as on average (relative abundances). That is a very complicated task given the limited sensitivity of the 
shower observables to the primary mass and the strong group superposition due to shower-to-shower fluctuations. The designed algorithm allows to include 
different types of constraints coming from the hadronic model predictions.\\We tested the algorithm over samples of data generated with different relative abundances. 
In the ideal case of perfect measurement precision and negligible contaminations from other species with respect to that used to fit the data we achieved 
good classification performances, with error rates around 10\%, comparable to that found with supervised methods (i.e. neural networks) in \cite{RiggiNIMPreparation}.
Component abundances and means can be reconstructed with good accuracy too.\\The classification performances drop off considerably in 
a more complicated situation in which a significant contamination from other nuclear species is present together with additive data fluctuations due to the 
experimental resolution. In this situation the discrimination of
the data into three general groups (light-intermediate-heavy) is still feasible, with typical accuracy in the relative abundances $\sim$0.05 depending on 
the degree of group contaminations.\\
The algorithm, as it is, can be applied to more than two shower observables (i.e. signal rise time or asymmetries, muon production depth, \dots) in a very 
straightforward way. The application to real cosmic ray data is easily done to, as we demonstrated in the case of data sample II, eventually explicitly 
including the effect of the real experimental resolution in the Monte Carlo templates and simplifying the model by ignoring the correlations relative to variables
measured with different detectors.

\end{document}